\pdfoutput=1

\documentclass[
submission, Phys]{SciPost}
\hypersetup{
    colorlinks,
    linkcolor={red!50!black},
    citecolor={blue!50!black},
    urlcolor={blue!80!black}
}

\DeclareSymbolFont{usualmathcal}{OMS}{cmsy}{m}{n}
\DeclareSymbolFontAlphabet{\mathcal}{usualmathcal}

\usepackage{float} 
\usepackage{graphicx}
\usepackage{subcaption}  
\usepackage{mathtools}
\usepackage{bm,bbm}

\usepackage{amsmath,amssymb}
\usepackage{amscd}
\usepackage{color}
\usepackage{spverbatim}
\usepackage{hyperref}

\newcommand{\IM}{\text{Im}}

\newcommand{\IMI}{\text{i}}

\providecommand{\ad}{a^{\dagger}}

\providecommand{\Hph}{H_{\mathrm{ph}}}
\providecommand{\HC}{\mathcal{H}_{\mathrm{C}}}
\providecommand{\HD}{\mathcal{H}_{\mathrm{D}}}
\providecommand{\Ua}{\mathcal{U}_{\alpha}}
\providecommand{\funding}[1]{\paragraph{Funding information}#1}

\begin{document}

\begin{center}{\Large \textbf{
      A universal gauge structure in the photon content of light-matter systems
    }}
\end{center}

\begin{center}
  Arka Dutta\textsuperscript{$\star$}
  and Marcus Kollar      
\end{center}

\begin{center}
      Theoretical Physics III, %
      Center for Electronic Correlations and Magnetism, %
      \newline%
      Institute of Physics, University of Augsburg, %
      86135 Augsburg, Germany
      \\
    ${}^\star$ {\small \sf arka.dutta@uni-a.de}
\end{center}

\begin{center}
  \today{}
\end{center}

\begin{abstract}
  After truncating the matter Hilbert space of a single-mode
  light-matter model, gauge transformations leave the spectrum
  invariant whereas the ground-state photon number is gauge dependent,
  hence, its meaning remains disputed. We show that this gauge
  dependence obeys an exact algebraic law whenever the gauge
  transformation acts as a displacement of the field coupled to a
  matter operator $G$. Namely, the ground-state photon number is a
  quadratic polynomial in the gauge parameter $\lambda$,
  $\langle a^{\dagger} a\rangle_{\lambda}$ $=$ $\langle a^{\dagger} a\rangle_{0}$ $+$
  $C\lambda$ $+$ $Q\lambda^2$. Here the coefficients are the
  correlator $C=\langle G\,p\rangle$ of $G$ and the field momentum
  $p$, and $Q=\langle G^2\rangle$. Its minimum over all gauges is
  naturally gauge-invariant, and for the quantum Rabi model it is
  determined by the ground-state energy alone. We show that it is
  strictly positive for the quantum Rabi and Dicke models with level
  splitting, so that a residual photon content survives in every
  gauge. At weak coupling the gauge of minimal photon number is the
  Jaynes-Cummings gauge, and for the Dicke model it remains so at all
  couplings in the thermodynamic limit.
\end{abstract}

\vspace{10pt}
\noindent\rule{\textwidth}{1pt}
\tableofcontents\thispagestyle{fancy}
\noindent\rule{\textwidth}{1pt}

  \newpage
  \section{Introduction}
  \label{sec:intro}

  Cavity and circuit quantum electrodynamics now reach a regime in
  which the light-matter coupling is comparable to, or larger than,
  the bare frequencies of the light and the matter
  excitations~\cite{kockum_ultrastrong_2019,forn-diaz_ultrastrong_2019,le_boite_theoretical_2020,schlawin_cavity_2022}. In
  this ultrastrong-coupling regime the rotating-wave approximation
  fails, virtual excitations dress the ground state, and the photon
  number depends strongly on the coupling. The quantum Rabi model
  (QRM)~\cite{rabi_process_1936}, describing a two-level system
  coupled to one bosonic mode, is the simplest model for this physical
  situation. Because it is exactly
  solvable~\cite{swain_continued_1973,braak_integrability_2011,braak_generalized_2013,chen_exact_2011,xieQuantumRabiModel2017}, it
  is also suitable for the study of conceptual questions that arise in
  this context.

  One conceptual aspect of the QRM and related models is the gauge
  ambiguity after the matter space is
  truncated~\cite{de_bernardis_breakdown_2018,stokes_gauge_2019,di_stefano_resolution_2019}. The
  standard QRM is written in the Coulomb gauge and is no longer
  gauge consistent after this truncation, whereas the truncated
  dipole-gauge QRM remains accurate.
  The reason for this gauge dependence is that a truncated potential
  is no longer local. Projecting onto two energy eigenstates gives the
  potential a momentum dependence,
  $V(x)\rightarrow\mathcal{P}V(x)\mathcal{P}=V(x,p)$, so that minimal
  coupling must act on it as well. The truncated
  Coulomb-gauge Hamiltonian therefore does not describe a
  gauge-symmetric theory. A robust remedy is to apply the gauge
  transformation \emph{before} truncating, which keeps the reduced
  Hamiltonian gauge consistent. We use this gauge-corrected
  Coulomb-gauge Hamiltonian
  throughout~\cite{di_stefano_resolution_2019}.

  Various gauge-related topics have recently been investigated.  For
  instance, it is still debated whether and how Jaynes-Cummings
  physics survives in the ultrastrong
  regime~\cite{stokes_gauge_2019,garziano_gauge_2020,savasta_thomasreichekuhn_2021,stokes_implications_2022,stokes_gauge_2023,stefano_reply_2023}.
  Moreover, an optimal gauge has been proposed that minimizes the
  truncation error of the ground-state energy or of dynamical
  correlators~\cite{qing_optimal_2025}. Also the ground state of the
  gauge-invariant Dicke model has been found to contain photons in
  squeezed, nonclassical states with no coherent order
  parameter~\cite{san_ground_2024}.

  Here we will address the question of how the photon number changes
  as one moves across a family of gauges.  The corrected Coulomb,
  dipole and general intermediate gauge
  Hamiltonians~\cite{di_stefano_resolution_2019} are all related by
  unitary transformations. However the ordinary unitary map between
  the \emph{uncorrected} Coulomb gauge and the others is not a gauge
  transformation, which is behind the apparent contradictions between
  the different formulations of the
  QRM~\cite{de_bernardis_breakdown_2018,dmytruk_gauge_2021}. Spectra
  are the same in every gauge and therefore cannot tell the gauges
  apart, while the photon-number operator does depend on the gauge.

  Our starting point is an exact law for this gauge dependence. Suppose
  the gauge change displaces the field by an amount that depends on
  the state of the matter. Then $\langle\ad a\rangle_{\lambda}$ $=$
  $\langle\ad a\rangle_{0}$ $+$ $C\lambda$ $+$ $Q\lambda^2$ holds
  exactly. Here $\lambda$ is the gauge parameter, $\lambda=0$ labels a
  reference gauge, and $G$ is the matter operator that couples to the
  field. The coefficient $C=\langle G\,p\rangle$ is the ground-state
  correlation between $G$ and the field momentum $p$ (defined below),
  and $Q=\langle G^2\rangle$ is the ground-state fluctuation of
  $G$. This is an algebraic identity and does not depend on details of
  the model or on whether the model is solvable. The physics lies in
  the minimum of the parabola, which is the same whichever gauge one
  starts from. For the QRM it is determined by the ground-state
  energy alone, so that the gauge-invariant part of the photon number
  is contained in the spectrum. The minimum cannot be
  negative, and we show in
  Secs.~\ref{subsec:photon-gauge} and~\ref{sec:dicke} that it
  is strictly positive for the Rabi and Dicke models with level
  splitting, so that the
  photon content it describes cannot be removed by any choice of
  gauge. In all cases that we examined, the gauge that reaches the
  minimum lies strictly between the Coulomb and dipole gauges. At weak
  coupling it is the Jaynes-Cummings gauge of
  Ref.~\cite{stokes_gauge_2019}, in which the counter-rotating terms
  vanish to linear order, and at very strong coupling it approaches
  the Coulomb gauge.

  In the Dicke model~\cite{dicke_coherence_1954,kirton_introduction_2019} the curvature is the
  collective dipole fluctuation, and its size is set by the dipole
  self-energy, i.e., the $P^2$ term of the dipole
  gauge~\cite{jaako_ultrastrong-coupling_2016}, which
  corresponds to the diamagnetic term at its Thomas-Reiche-Kuhn (TRK)
  value in the Coulomb gauge~\cite{garziano_gauge_2020}. This term
  makes the gauge family
  unitarily consistent and rules out the superradiant phase
  transition~\cite{hepp_superradiant_1973,rzazewski_phase_1975,nataf_nogo_2010}. Gauge
  invariance and the no-go theorem therefore come from the same
  term. Without it the model shows the familiar superradiant
  instability, which is an artifact of dropping this term, the
  curvature grows with coupling, and the minimal photon number turns
  over at a coupling that approaches the critical one as the number
  of emitters grows. With the self-energy restored, none of this
  happens, and in the thermodynamic limit the gauge of minimal photon
  number is the Jaynes-Cummings gauge at all couplings.

  The minimum photon number has an operational meaning, as it can be
  obtained from a conditional measurement protocol: for the QRM
  it equals the photon number of the field conditioned on a
  measurement of the dipole, after subtraction of the conditional mean
  field, and for the Dicke model it is bounded from below by this
  quantity. A direct measurement of the field noise instead yields a
  larger value, which exceeds the minimum by an amount set by a
  light-matter correlator. We discuss such measurements in
  Sec.~\ref{subsec:operational} and at the end of
  Sec.~\ref{sec:dicke}.

\begin{figure}[t]
    \centering
    \includegraphics[width=\linewidth]{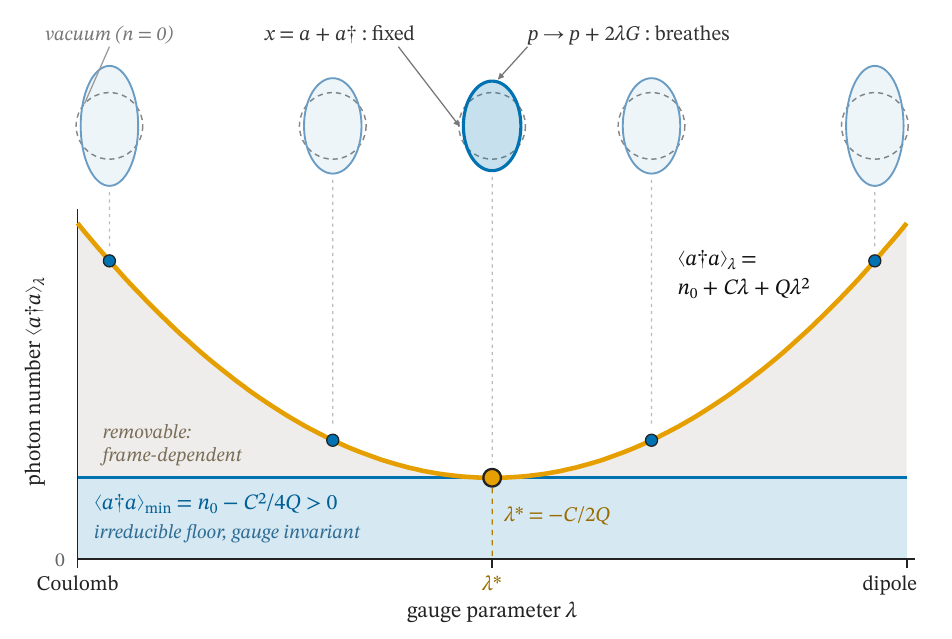}
    \caption{The gauge family of the photon dressing. Changing the
      gauge parameter $\lambda$ in
      $U_{\lambda}=\exp[i\lambda G(a+\ad)]$ leaves the position
      quadrature $x$ of the field unchanged and both shifts and
      reshapes the momentum quadrature $p$. The top row shows
      phase-space insets from the Coulomb gauge (left), through
      $\lambda_{\ast}$, to the dipole gauge (right). The momentum
      fluctuation $\langle p^2\rangle_{\lambda}$ is itself a
      parabola in $\lambda$, narrowest at $\lambda_{\ast}$. The
      ground-state photon number $\langle\ad a\rangle_{\lambda}$
      follows the matching exact parabola (bottom), and its minimum at
      $\lambda_{\ast}$ is strictly positive and the same in every
      gauge, corresponding to the photon dressing that no gauge choice
      removes (Sec.~\ref{sec:universal}).}
    \label{fig:gauge-orbit-overview}
  \end{figure}

  Figure~\ref{fig:gauge-orbit-overview} gives an overview. The gauge
  unitary $U_{\lambda}=\exp[i\lambda G(a+\ad)]$ leaves the position
  quadrature $x=a+\ad$ of the field unchanged and shifts the momentum
  quadrature $p=i(\ad-a)$ by $2\lambda G$. Because $G$ is correlated
  with the field in the entangled ground state, the momentum
  fluctuation changes as well:
  $\langle p^2\rangle_{\lambda}=\langle p^2\rangle_{0}+4\lambda
  C+4\lambda^2Q$, a parabola for which the minimum lies at the same
  value $\lambda_{\ast}$ of the gauge parameter as that of the photon
  number. Going from the Coulomb to the dipole gauge therefore moves
  and reshapes the field distribution along one quadrature, as the
  insets show, and this traces out the gauge orbit of the photon
  dressing.

  The paper is organized as follows. Section~\ref{sec:universal}
  derives the gauge parabola in general. For the QRM, we recall the
  Coulomb, dipole and general $\alpha$-gauge Hamiltonians in
  Sec.~\ref{sec:gauge}, showing how they are connected by unitary
  transformations, and map the arbitrary-gauge model onto an effective
  one-photon standard QRM.  Section~\ref{subsec:coh-equivalence}
  solves this model with the method of tunable coherent states, which
  is equivalent to the $G$-function
  method~\cite{braak_integrability_2011} and the confluent-Heun
  representation~\cite{zhong_analytical_2013}. In
  Secs.~\ref{subsec:photon-standard} and~\ref{subsec:photon-gauge} we
  determine the photon number across the entire gauge family in closed
  form and discuss how it could be measured
  (Sec.~\ref{subsec:operational}). Sec.~\ref{sec:dicke} examines the
  Dicke model, where exact diagonalization confirms the parabola and
  links its curvature to the dipole self-energy, and through it to
  gauge invariance and the no-go theorem for superradiance, and where
  the thermodynamic limit can be obtained
  analytically. Sec.~\ref{sec:conclusion} concludes with an outlook.

  \section{The universal parabolic gauge structure}
  \label{sec:universal}
  \label{subsec:universal-parabola}

  Once the matter Hamiltonian is truncated to a finite number of
  energy levels, any single-mode problem like the QRM has a
  one-parameter family of unitarily equivalent gauges. They all have
  the same spectrum but differ in their ground-state photon
  number. We first show that this photon number has a fixed
  form for all such models, whenever the gauge change displaces the
  field by an amount that depends on the state of the matter. Consider
  any single-mode light-matter Hamiltonian for which a gauge
  transformation is generated by
  \begin{align}
    U_{\lambda} = \exp\!\bigl[i\,\lambda\,G\,(a+\ad)\bigr],
    \label{eq:universal-U}
  \end{align}
  where $G=G^{\dagger}$ is the Hermitian matter operator that
  multiplies the field quadrature $a+\ad$ in the interaction term of
  the Hamiltonian, and $\lambda$ is a real parameter that labels the
  gauge. Different values of $\lambda$ pick out the Coulomb, dipole
  and intermediate gauges of a given model, as detailed for the QRM in
  Sec.~\ref{subsec:mapping} and for the Dicke model in
  Sec.~\ref{sec:dicke}.
  
  Because $G$ acts only on the matter, it commutes with $a$ and $\ad$,
  and therefore also with the generator $X\equiv i\lambda G(a+\ad)$.
  We write $U_{\lambda}^{\dagger}\,a\,U_{\lambda}=e^{-X}a\,e^{X}$ and
  expand in nested commutators. The Baker-Campbell-Hausdorff series
  stops after the first term, since $[X,a]$ $=$ $-\IMI\lambda\,G$ and
  $[X,-\IMI\lambda G]$ $=$ $0$. Therefore
  \begin{align}
    U_{\lambda}^{\dagger}\,a\,U_{\lambda}=a+i\lambda G
    \label{eq:universal-displacement}
  \end{align}
  holds exactly, to all orders in $\lambda$.
  The photon-number operator then transforms as
  \begin{align}
    U_{\lambda}^{\dagger}\,\ad a\,U_{\lambda} = \ad a + i\lambda\,G\,(\ad-a) + \lambda^2\,G^2.
    \label{eq:universal-transform}
  \end{align}
  Taking the ground-state expectation value gives, for any such model,
  \begin{align}
    \langle\ad a\rangle_{\lambda} = \langle\ad a\rangle_{0} + C\,\lambda + Q\,\lambda^2,
    \qquad
    C=i\langle G(\ad-a)\rangle_{0},\quad Q=\langle G^2\rangle_{0}.
    \label{eq:universal-parabola}
  \end{align}
  The photon number is therefore an exact parabola in $\lambda$ with
  the ground state fluctuation $Q$ as its curvature. The minimum
  photon number among all gauges is thus
  \begin{align}
    \langle\ad a\rangle_{\min}
    &=
    \langle\ad a\rangle_{\lambda_\ast}
      =
      \langle\ad a\rangle_{0}-\frac{C^2}{4Q}
      \,,
    &
      \lambda_{\ast}
    &=
      -\frac{C}{2Q}
      \,,
      \label{eq:universal-minphotons}
  \end{align}
  and does not depend on how the gauge family is parametrized.  In
  particular, the QRM corresponds to $G=\sigma_{x}$, so that $G^2$ is
  the identity and the curvature is exactly one, as it is for all
  two-level matter systems.  We note that a gauge with no photons at
  all can exist only if the discriminant
  $C^{2}-4Q\langle\ad a\rangle_{0}$ vanishes. It can never be
  positive, as follows from
  $C=2\,\IM\langle G\psi_{0}|a\psi_{0}\rangle$ and the Cauchy-Schwarz
  inequality,
  \begin{align}
    Q\,\langle\ad a\rangle_{\min}
    =\|G\psi_{0}\|^2\|a\psi_{0}\|^2-\bigl(\IM\langle G\psi_{0}|a\psi_{0}\rangle\bigr)^2
    \geq\|G\psi_{0}\|^2\|a\psi_{0}\|^2-|\langle G\psi_{0}|a\psi_{0}\rangle|^2
    \geq0,
    \label{eq:cauchy-schwarz}
  \end{align}
  where $|\psi_{0}\rangle$ is the ground state in the reference
  gauge. For the Rabi and Dicke models an antiunitary symmetry makes
  $\langle Ga\rangle_{0}$ purely imaginary
  (cf.\ Sec.~\ref{subsec:operational} and
  Ref.~\cite{braak_symmetries_2019}), so that the first inequality
  is an equality and $\langle\ad a\rangle_{\min}$ is the squared norm
  of the part of $a|\psi_{0}\rangle$ that is orthogonal to
  $G|\psi_{0}\rangle$. The minimum vanishes if and only if
  $(a+i\lambda G)|\psi_{0}\rangle=0$ for some $\lambda$, i.e., if and
  only if in some gauge the ground state factorizes with the field in
  its vacuum,
  $U_{\lambda}|\psi_{0}\rangle=|0\rangle\otimes|\chi\rangle$. For the
  QRM and the Dicke model with level splitting this is impossible, as
  we show in Secs.~\ref{subsec:photon-gauge} and~\ref{sec:dicke}, so
  that $\langle\ad a\rangle_{\min}>0$.

  For both cases the gauge that minimizes the photon number lies in
  between Coulomb gauge ($\alpha=0$) and dipole gauge ($\alpha=1$).
  For the QRM we find $\alpha_{\min}=-\lambda_{\ast}/\eta$, where
  $\eta=g/\omega_{c}$ is the normalized coupling and $\alpha$ is
  defined in~\eqref{eq:alpha-def}. For the Dicke model
  $\alpha_{\min}=-\lambda_{\ast}/g$, with $g=\eta\sqrt{N_{s}}$ the
  collective coupling
  (see Secs.~\ref{sec:gauge} and~\ref{sec:dicke} for
  definitions). We find numerically that $0<\alpha_{\min}<1$ in every
  case, as illustrated in Figs.~\ref{fig:alpha-min}
  and~\ref{fig:dicke-alpha-min}. For the QRM the upper bound
  $\alpha_{\min}<1$ can also be shown in general
  (Sec.~\ref{subsec:photon-gauge}).
  
  We establish this structure in two settings. For the QRM
  (Sec.~\ref{sec:gauge}) we recall the gauge
  family, map the standard gauge to an effective one-photon model, and
  solve it with tunable coherent states, without a
  photon-number cutoff. The resulting parabola, with $G=\sigma_{x}$
  and unit curvature, is then evaluated across the entire gauge
  family in closed form by composing the standard-gauge
  solution with the two exactly solvable maps that reach the
  gauge-corrected Coulomb gauge.  For the Dicke model
  (Sec.~\ref{sec:dicke}), where
  $G=N_{s}^{-1/2}\sum_{l}\sigma^{x}_{l}=2N_{s}^{-1/2}\hat J_{x}$, with
  the collective spin $\hat J_{x}=\tfrac12\sum_{l}\sigma^{x}_{l}$,
  exact diagonalization confirms the parabola for $N_{s}$ emitters and
  identifies the curvature $Q=\langle G^2\rangle$, which is set by the
  dipole self-energy, the term that makes the gauges unitarily
  equivalent and enforces the no-go theorem for superradiance.

  We remark that the generator $G(a+\ad)$ is not an arbitrary choice,
  but rather encapsulates the full gauge freedom of this class of
  models.  The gauge function is the product of the matter dipole and
  the vector potential, which after truncation is $G$ times the field
  quadrature $x$ or $p$ to which $A$ is proportional.  Which
  quadrature stays fixed thus depends only on with which the matter
  operator $G$ multiplies in the coupling. In this work the latter is
  $x$, because every Hamiltonian we study couples through
  $(a+\ad)$. For completeness we note that the same structure appears
  for the conjugate quadrature, although this is not a gauge
  transformation of the Hamiltonians studied here. The generator built
  on $p$, $U_{\lambda}'=\exp[i\lambda Gp]=\exp[\lambda G(a-\ad)]$, is
  also unitary, since $G(a-\ad)$ is anti-Hermitian. The calculation
  behind~\eqref{eq:universal-displacement} then gives
  $U_{\lambda}'^{\dagger}pU_{\lambda}'=p$ and
  $U_{\lambda}'^{\dagger}xU_{\lambda}'=x-2\lambda G$. The two
  quadratures simply swap roles: $p$ stays fixed, $x$ is displaced,
  and its variance follows a parabola,
  $\langle x^2\rangle_{\lambda}=\langle x^2\rangle_{0}-4\lambda\tilde
  C+4\lambda^2Q$, with $\tilde C=\langle Gx\rangle_{0}$ in place of
  $C=\langle Gp\rangle_{0}$. Only $Q=\langle G^2\rangle$, and with it
  the photon-number parabola~\eqref{eq:universal-parabola}, is the
  same for both choices, because $\ad a$ is built symmetrically from
  the two quadratures. Hence only the coupling determines which
  quadrature plays which role. We also note that for several modes,
  or beyond the dipole approximation, the same structure holds as long
  as the matter operators $G_{k}$ that couple to the different modes
  commute with each other (which holds before, but not necessarily
  after the truncation), and the curvature then becomes a sum over the
  modes.

  \section{The gauge-correct~QRM and its relation to the standard QRM}
  \label{sec:gauge}
  
  \subsection{Coulomb, dipole and $\alpha$ gauges}
  \label{subsec:gauges}
  
  We first recall the general gauge setup for the QRM that was
  established
  in~\cite{de_bernardis_breakdown_2018,di_stefano_resolution_2019}. For
  a single quantum particle coupled to a single cavity mode in the
  electric-dipole approximation, the Coulomb-gauge Hamiltonian is
  \begin{align}
    H_{\mathrm{C}} = \frac{1}{2m}\bigl(p - q\hat{A}\bigr)^2 + V(x) + \Hph ,
    \label{eq:HC-particle}
  \end{align}
  where $\hat{A} = A_{0}(a+\ad)$ is the vector potential at the
  position of the particle, with zero-point amplitude $A_{0}$, and
  $\Hph = \omega_{c}\,\ad a$ is the Hamiltonian of the cavity field.
  Projecting~\eqref{eq:HC-particle} onto a two-level matter space
  gives~\cite{de_bernardis_breakdown_2018,di_stefano_resolution_2019}
  \begin{align}
    \mathcal{H}_{\mathrm{C}}'
    = \Hph + \Delta\sigma_{z} + g\,\sigma_{y}\,(\ad+a)
    + D\,(\ad+a)^2,
  \end{align}
  with $g=\omega_{10}A_{0}d_{10}$, $D=q^2A_{0}^2/(2m)$ and dipole
  matrix element $d_{10}\equiv q\langle 1|x|0\rangle$. The $(\ad+a)^2$
  term can be removed by the Bogoliubov
  transformation~\cite{yoshihara_superconducting_2017},
  \begin{align}
    b &= \tfrac{1}{2}\bigl[(\kappa+\kappa^{-1})a + (\kappa-\kappa^{-1})\ad\bigr],
        &
    b^{\dagger} &= \tfrac{1}{2}\bigl[(\kappa-\kappa^{-1})a + (\kappa+\kappa^{-1})\ad\bigr],
  \end{align}
  with $\tilde{\omega}_{c}=\sqrt{\omega_{c}^2+4\omega_{c}D}$ and
  $\kappa=\sqrt{\tilde{\omega}_{c}/\omega_{c}}$, maintaining
  $[b,b^{\dagger}]=1$. After a $\pi/2$ rotation about the $z$ axis the
  coupling is linear, renaming $b$ $\to$ $a$, and omitting tildes
  gives the \emph{standard} QRM Hamiltonian
  \begin{align}
    H_{\mathrm{std}} = \omega_{c}\,\ad a + g\,(a+\ad)\,\sigma_{x} + \Delta\,\sigma_{z}.
    \label{eq:standard-QRM}
  \end{align}
  The dipole-gauge Hamiltonian $\mathcal{H}_{\mathrm{D}}$, on the
  other hand, is the Power-Zienau-Woolley Hamiltonian in the dipole
  approximation~\cite{babiker_derivation_1983,woolley_electrodynamics_1975}. It
  reads
  \begin{align}
    \mathcal{H}_{\mathrm{D}} = \Hph + i\,g_{\mathrm{D}}\,(\ad - a)\,\sigma_{x} + \Delta\,\sigma_{z}
    + \eta^2\omega_{c}
    \,,
    \label{eq:dipole-QRM}
  \end{align}
  where the constant dipole self-energy $\eta^2\omega_{c}$ (with
  $\eta$ defined below) is kept for consistency with the gauge family.
  It can be safely gauge-transformed by means of the
  unitary~\cite{di_stefano_resolution_2019}
  \begin{align}
    \mathcal{U} = \exp\!\bigl[-\IMI\,\eta\,\sigma_{x}\,(a+\ad)\bigr],
    \label{eq:U-correction}
  \end{align}
  leading to the Hamiltonian in the corrected Coulomb gauge,
  \begin{align}
    \HC = \omega_{c}\, \ad a
    + \Delta\left\{\sigma_{z}\cos\left[2\eta(a+\ad)\right]
    + \sigma_{y}\sin\left[2\eta(a+\ad)\right]\right\},
    \label{eq:corrected-Coulomb}
  \end{align}
  which, compared to the standard QRM~\eqref{eq:standard-QRM},
  contains nonlinear corrections in field operators.  Here
  $\eta = g_{\mathrm{D}}/\omega_{c}$ is the normalized
  coupling.  The gauge freedom can be labeled by a real parameter
  $\alpha$, through the function operator
  $X_{\alpha} = \alpha\,q\,x\,\hat{A}$, which generates the gauge
  transformation $\Ua = \exp[-\IMI X_{\alpha}]$ in the dipole
  approximation. The values $\alpha = 0$ and $\alpha = 1$ give the
  Coulomb and dipole gauges. After projection onto two levels the
  unitary becomes $\Ua = \exp[-\IMI\alpha\,\eta\,\sigma_{x}(a+\ad)]$,
  and the corresponding intermediate $\alpha$-gauge Hamiltonian is
  \begin{align}
    \mathcal{H}^{(\alpha)} &= \Ua\,\HC\,\Ua^{\dagger},
                             \nonumber\\
    &= \omega_{c}\,\ad a
                                - \IMI\,\alpha\,g_{\mathrm{D}}\,(a-\ad)\,\sigma_{x}
                                + \alpha^2\eta^2\omega_{c} \nonumber\\
                              &~~~~~~+ \Delta\Bigl\{\sigma_{z}\cos\!\bigl[2\eta(1-\alpha)(a+\ad)\bigr]
                                + \sigma_{y}\sin\!\bigl[2\eta(1-\alpha)(a+\ad)\bigr]\Bigr\}.
    \label{eq:alpha-def}
  \end{align}
  This family of gauge-correct QRM Hamiltonians will be studied in the
  following, and next we will show how to map it onto a standard QRM.
  Throughout we assume a level splitting $\Delta>0$ and a nonzero
  coupling.

  \subsection{Mapping to the standard QRM}
  \label{subsec:mapping}
  
  We now map the $\alpha$-gauge QRM Hamiltonian~\eqref{eq:alpha-def}
  to a standard QRM in order to benefit from the available exact
  solution methods for the latter.  The standard
  QRM~\eqref{eq:standard-QRM} is mapped to the dipole gauge
  Hamiltonian, up to the constant $\eta^2\omega_{c}$, by the unitary
  transformation~\cite{de_bernardis_breakdown_2018}
  \begin{align}
    U = \exp\!\left(\IMI\,\tfrac{\pi}{2}\,\ad a\right),
    \label{eq:rotation-CtoD}
  \end{align}
  which acts as $a\rightarrow -\IMI a$, $\ad\rightarrow \IMI\ad$ and
  relates the couplings by $g_{\mathrm{D}} = g_{\mathrm{std}}$.
  Importantly, the unitary~\eqref{eq:rotation-CtoD} is \emph{not} a
  gauge transformation. However it allows us to map the standard QRM
  to the entire $\alpha$-gauge family as summarized in the following
  diagram,
  \begin{center}
    \resizebox{0.42\textwidth}{!}{$%
      \begin{CD}
        \mathcal{H}_{\mathrm{D}}
        @<{U=\exp[i\frac{\pi}{2}\ad a]}<<
        H_{\mathrm{std}} \\
        @V{\mathcal{U}=\exp[-i\eta\sigma_{x}(a+\ad)]}VV @.\\
        \HC
        @>{\Ua=\exp[-i\alpha\eta\sigma_{x}(a+\ad)]}>>
        \mathcal{H}^{(\alpha)}
      \end{CD}$}
  \end{center}
  where the horizontal arrows act on the bosonic mode and the vertical
  switches between dipole and corrected Coulomb gauge. The directions
  of the arrows indicate how an eigenstate of $H_{\mathrm{std}}$ maps
  onto an eigenstate of $\mathcal{H}^{(\alpha)}$ with the same energy.
  We emphasize that $H_{\mathrm{std}}$ serves here only as a
  convenient representation of $\HD$ for which exact solutions are
  available; its eigenstates are not those of a physical Coulomb-gauge
  model.

  \subsection{Exact solution of the one-photon QRM via tunable coherent states}
  \label{subsec:coh-equivalence}

  We make a tunable coherent
  states ansatz~\cite{xie_exact_2019,li_two-photon_2020}, which
  builds on the extended coherent states introduced for
  polarons~\cite{qinghu_extensive_1994,qinghu_two-_1996} and for the
  QRM~\cite{chen_exact_2011}, and in which
  the two components refer to the eigenbasis of $\sigma_{x}$, which is
  coupled by the $\Delta$ term:
  \begin{align}
    |\Psi\rangle =
    \binom{\sum_{n=0}^{\infty} c_{n}\,(\ad)^{n}\exp(\xi\ad)\,|0\rangle}
    {\Pi\sum_{n=0}^{\infty} c_{n}\,(-\ad)^{n}\exp(-\xi\ad)\,|0\rangle},
    \label{eq:tunable-ansatz}
  \end{align}
  with coefficients $c_{n}$ and a coherent displacement $\xi$,
  which clearly encompasses all possible
  states. Substituting~\eqref{eq:tunable-ansatz} into the
  Schr\"odinger equation for~\eqref{eq:standard-QRM} and matching the
  coefficients of $(\ad)^{n}\exp(\xi\ad)|0\rangle$ at order $n$ gives
  the recursion~\cite{xie_exact_2019}
  \begin{align}
  E\,c_{n} = (n+\xi g)\,c_{n} + (\xi+g)\,c_{n-1} + g(n+1)\,c_{n+1}
    - (-1)^{n}\,\Pi\Delta\sum_{j=0}^{n}\frac{(2\xi)^{j}}{j!}\,c_{n-j}.
    \label{eq:tunable-recursion}
  \end{align}
  We set $c_{0}=1$ as normalization, and as the displacement $\xi$
  already carries the term linear in $\ad$ one can also take
  $c_{1}=0$. This lets us express the energy as
  \begin{align}
    E = \xi g - \Pi\Delta,
    \label{eq:tunable-energy}
  \end{align}
  and putting~\eqref{eq:tunable-energy} back
  into~\eqref{eq:tunable-recursion} yields a closed recursion
  \begin{align}
    c_{n+1} = \frac{(n+\Pi\Delta)\,c_{n} + (\xi+g)\,c_{n-1}}{-g(n+1)}
    + \frac{\Pi\Delta\,(-1)^{n}}{g(n+1)}\sum_{j=0}^{n}\frac{(2\xi)^{j}}{j!}\,c_{n-j}.
    \label{eq:tunable-coefficient-recursion}
  \end{align}
  Finally, normalizability of the state requires the $c_n$ to vanish
  at infinity sufficiently fast.
  
  We note that this construction is equivalent to the $G$-function
  method~\cite{braak_integrability_2011} and confluent-Heun
  representation~\cite{zhong_analytical_2013}, for the following
  reason. Define the generating function $c(z)=\sum_{n}c_{n}z^{n}$ and
  multiply~\eqref{eq:tunable-coefficient-recursion} by $z^{n}$ and sum
  over $n$, using $\sum_{n}(n+1)c_{n+1}z^{n}$ $=$ $c'(z)$,
  $\sum_{n}c_{n-1}z^{n}$ $=$ $z\,c(z)$ and
  $\sum_{n}\sum_{j}(-1)^{n}\frac{(2\xi)^{j}}{j!}c_{n-j}z^{n}$ $=$
  $c(-z)e^{-2\xi z}-1$. The substitution $c(z)$ $=$
  $\phi_{1}(z)e^{-\xi z}$ then gives exactly the even-parity equation
  for $\phi_{1}(z)$~\cite{braak_integrability_2011} with $E$ $=$
  $\xi g-\Delta$. The odd-parity sector follows from $c(z)$ $=$
  $\phi_{2}(z)e^{\xi z}$ with $E=\xi g+\Delta$, together
  recovering~\eqref{eq:tunable-energy}.  Hence the tunable
  coefficients $c(z)$ can be written in terms of confluent Heun
  functions, and all three approaches describe the same
  spectrum. The tunable coherent state ansatz also provides the
  explicit eigenstate, which will be used below for the photon number
  and the correlator $C^{\mathrm{std}}$, although the minimum photon
  number can be obtained from the eigenenergy alone
  (Sec.~\ref{subsec:photon-gauge}).

\begin{figure}[t]
    \centering
    \includegraphics[width=\linewidth]{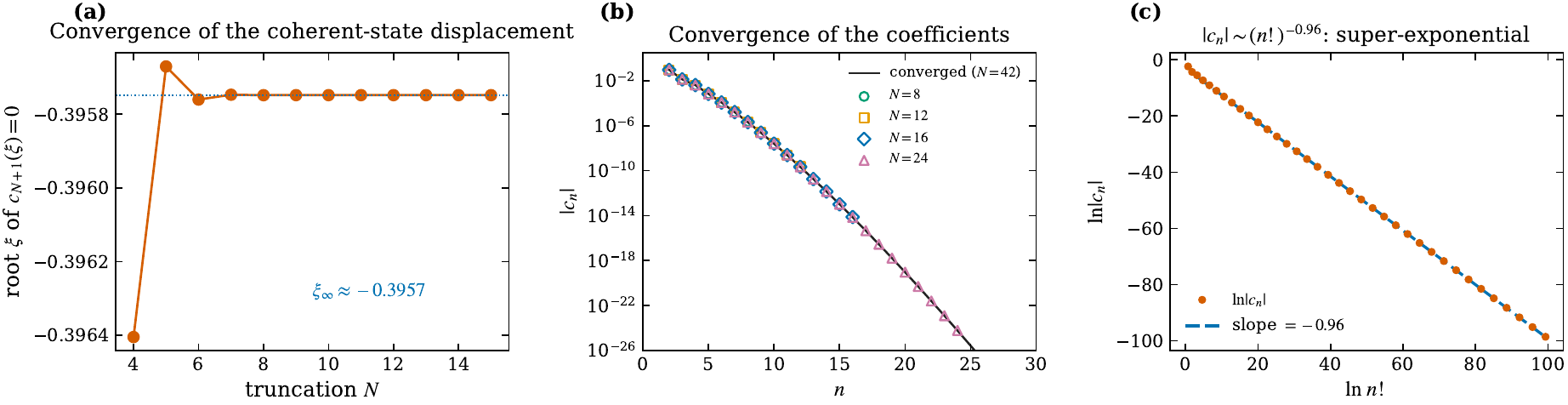}
    \caption{Validity of the truncated tunable-coherent-state
      expansion for $\Delta=0.7$, $\Pi=1$, $g=0.8$. (a) Convergence of
      the coherent-state parameter $\xi$ with the truncation $N$. (b)
      Coefficient magnitudes $|c_n|$ for $N=8,12,16,24$ (markers,
      shown only up to $n=N$), which fall on the converged curve
      (line) and reach about $10^{-25}$ at $N=24$. (c) $\ln|c_n|$
      against $\ln n!$ is linear with slope $\approx-0.96$, so
      $|c_n|\sim(n!)^{-0.96}$, a decay faster than any geometric
      rate. The displacement $\xi$ obtained from the truncation
      condition $c_{N+1}(\xi)=0$ converges quickly with $N$, so a
      short truncation already gives a numerically exact approximation
      of the Rabi wavefunctions.}
    \label{fig:tunable-validity}
  \end{figure}

  In practice, the solution in terms of an
  ansatz~\eqref{eq:tunable-ansatz} is determined from the displacement
  $\xi$ after fixing a truncation condition $c_{N+1}=0$ at finite $N$,
  yielding a polynomial equation in $\xi$, the relevant root of which
  (the one with the lowest energy~\eqref{eq:tunable-energy})
  converges quickly with $N$ (Fig.~\ref{fig:tunable-validity}(a)).
  The truncation condition thus acts as a quantization condition, the
  convergence of which we check numerically. The
  coefficients themselves decay faster than any geometric rate: at
  high precision we find $|c_{n}|\sim(n!)^{-0.96}$
  (Fig.~\ref{fig:tunable-validity}(b,c)), which presumably reflects
  $|c_{n}|\sim\tau^{n}/n!$ with power-law corrections, as expected for
  a generating function $c(z)$ of exponential type. The coherent factor
  $e^{\xi\ad}$ already carries most of the Fock-space weight, so a
  rather short truncation is enough. While exact diagonalization has
  to cut the Fock space off at a finite photon number, the present
  method truncates only the expansion in displaced Fock states, so
  the solution is numerically exact at any coupling strength.

  \subsection{Photon number in the standard gauge}
  \label{subsec:photon-standard}

  With the eigenstates~\eqref{eq:tunable-ansatz} in hand, the
  ground-state photon number can be determined as follows. After
  normalizing the state, the expectation value
  $\langle\ad a\rangle$ involves matrix elements
  $I_{i,j}(\xi)=\langle 0|e^{\xi
    a}a^{j}(\ad)^{i}e^{\xi\ad}|0\rangle$. They are generated by
  \begin{align}
    \sum_{n,m=0}^{\infty}\frac{X^{n}Y^{m}}{n!\,m!}\,I_{m,n}(\xi)
    = \langle 0|e^{(\xi+X)a}e^{(\xi+Y)\ad}|0\rangle
    = e^{\xi^2+\xi(X+Y)+XY},
    \label{eq:generating}
  \end{align}
  so that $I_{i,j}(\xi)=e^{\xi^2}P_{i,j}(\xi)$, where the
  $P_{i,j}(\xi)$ are two-variable Hermite-like polynomials that obey
  \begin{align}
    P_{i+1,j}(\xi) &= \xi P_{i,j}(\xi) + j\,P_{i,j-1}(\xi), \nonumber\\
    P_{i,j+1}(\xi) &= \xi P_{i,j}(\xi) + i\,P_{i-1,j}(\xi),
                     \label{eq:P-recursion}
  \end{align}
  with $P_{0,0}=1$, $P_{1,0}=P_{0,1}=\xi$ and
  $P_{1,1}=\xi^2+1$. Collecting terms gives a compact expression for
  the ground-state photon number,
  \begin{align}
    \langle\ad a\rangle
    = \frac{\sum_{i,j}c_{i}c_{j}\,P_{i+1,j+1}(\xi)}
    {\sum_{i,j}c_{i}c_{j}\,P_{i,j}(\xi)} - 1.
    \label{eq:photon-standard}
  \end{align}
  We evaluate this expression by truncating $i,j$ at the same cutoff
  $N$ as the coefficients. The same construction gives the
  standard-gauge correlator
  $C^{\mathrm{std}}=\langle\sigma_{x}(a+\ad)\rangle_{0}$, which we
  need in Sec.~\ref{subsec:photon-gauge},
  \begin{align}
    C^{\mathrm{std}} = \frac{\sum_{i,j}c_{i}c_{j}\bigl(P_{i,j+1}(\xi)+P_{i+1,j}(\xi)\bigr)}
    {\sum_{i,j}c_{i}c_{j}\,P_{i,j}(\xi)},
    \label{eq:C-std}
  \end{align}
  using  the same polynomials and coefficients as~\eqref{eq:photon-standard}.
  
\begin{figure}[t]
    \centering
    \includegraphics[width=0.78\linewidth]{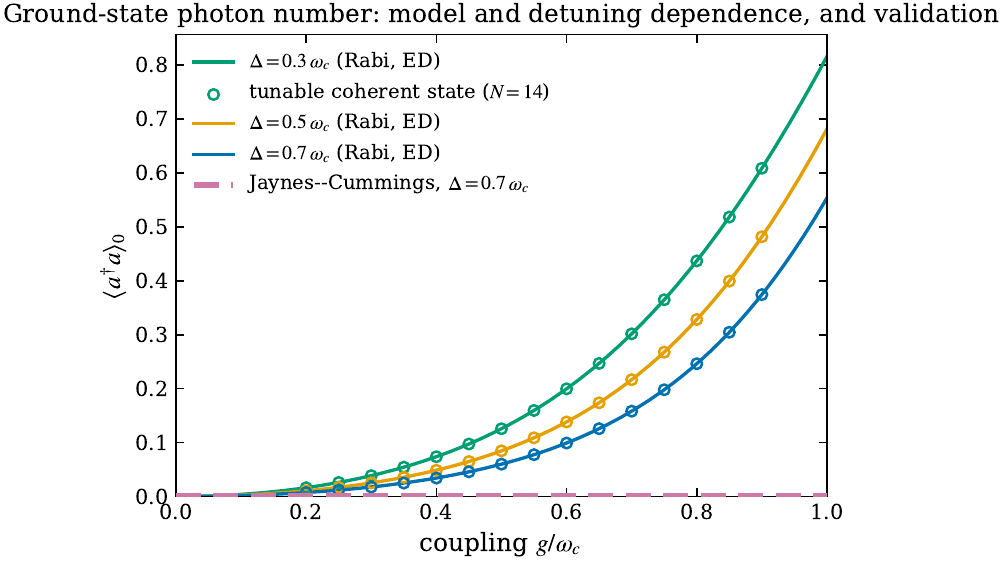}
    \caption{Ground-state photon number of the standard Rabi
      model~\eqref{eq:standard-QRM}, which equals that in the dipole
      gauge, versus coupling $g$, from exact diagonalization
      ($N_{\mathrm{cutoff}}=160$, $\omega_{c}=1$), for three detunings
      $\Delta=0.3,0.5,0.7\,\omega_{c}$. The curves grow monotonically
      as the counter-rotating terms dress the ground state. The
      Jaynes-Cummings counterpart at $\Delta=0.7\,\omega_{c}$ stays at
      zero, because its ground state decouples exactly from the
      coupling for $g<\sqrt{2\omega_{c}\Delta}\approx1.18\,\omega_{c}$.
      The two models therefore separate strongly once the
      counter-rotating terms matter, which is where the gauge question
      becomes consequential. Open circles show the
      tunable-coherent-state result ($N=14$) at each of the three
      detunings. They reproduce the corresponding
      exact-diagonalization curve over the whole range of couplings.}
    \label{fig:photon-validation}
  \end{figure}

  We checked these analytic expressions against exact diagonalization. In
  the dressed basis the Hamiltonian acts as
  \begin{align}
    H_{R}|n,\uparrow\rangle &= (n\omega_{c}+\Delta)|n,\uparrow\rangle
                              + g\sqrt{n+1}\,|n+1,\downarrow\rangle + g\sqrt{n}\,|n-1,\downarrow\rangle, \nonumber\\
    H_{R}|n,\downarrow\rangle &= (n\omega_{c}-\Delta)|n,\downarrow\rangle
                                + g\sqrt{n+1}\,|n+1,\uparrow\rangle + g\sqrt{n}\,|n-1,\uparrow\rangle,
  \end{align}
  and the Hellmann-Feynman relation
  $\langle\ad a\rangle = \partial\langle H\rangle/\partial\omega_{c} =
  \partial E/\partial\omega_{c}$ gives the photon number from the
  numerical derivative of the ground-state energy. As
  Fig.~\ref{fig:photon-validation} shows, the tunable coherent-state
  result and the exact-diagonalization result agree essentially
  perfectly, the latter with photon cutoff
  $N_{\mathrm{cutoff}}=160$. Applying the same exact diagonalization
  to the Jaynes-Cummings model shows that the photon number vanishes,
  because for $g<\sqrt{2\omega_{c}\Delta}$ its ground state
  $|0,\downarrow\rangle$ decouples from the coupling.

  \subsection{Photon number across gauges}
  \label{subsec:photon-gauge}
  
\begin{figure}[t]
    \centering
    \begin{subfigure}{0.49\textwidth}
      \centering
      \includegraphics[width=\linewidth]{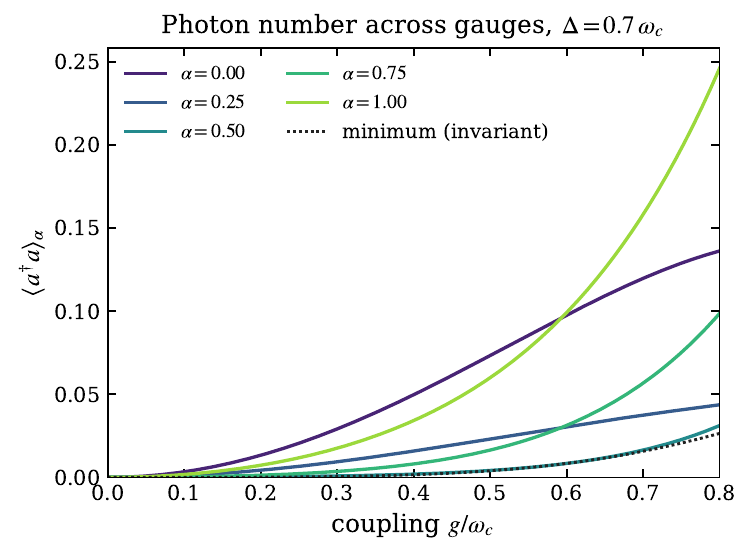}
      \caption{}
      \label{fig:photon-vs-g-alpha}
    \end{subfigure}\hfill
    \begin{subfigure}{0.49\textwidth}
      \centering
      \includegraphics[width=\linewidth]{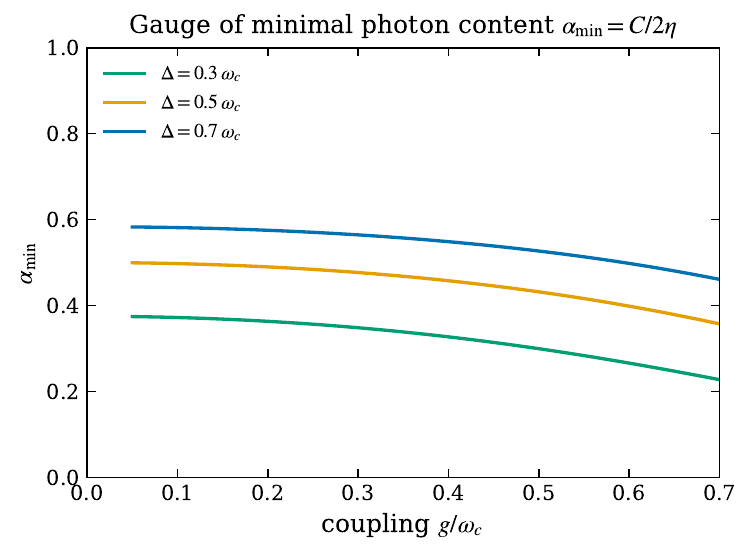}
      \caption{}
      \label{fig:alpha-min}
    \end{subfigure}
    
    \vspace{0.6em}
    
    \begin{subfigure}{0.49\textwidth}
      \centering
      \includegraphics[width=\linewidth]{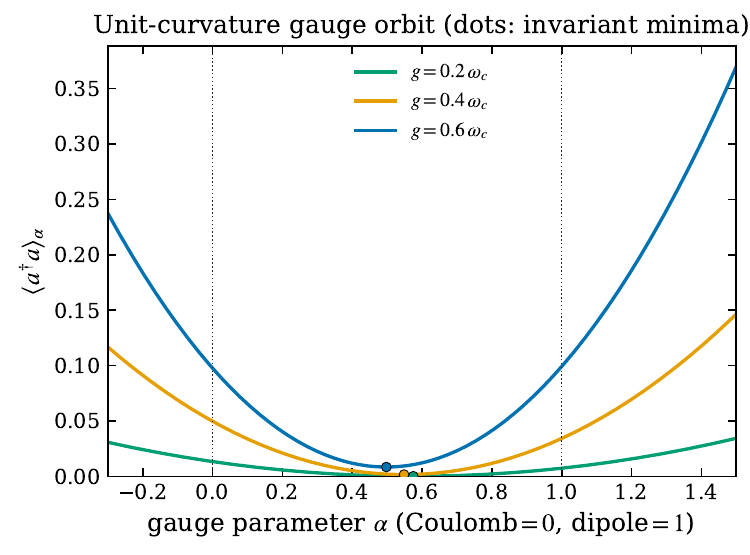}
      \caption{}
      \label{fig:photon-vs-alpha}
    \end{subfigure}\hfill
    \begin{subfigure}{0.49\textwidth}
      \centering
      \includegraphics[width=\linewidth]{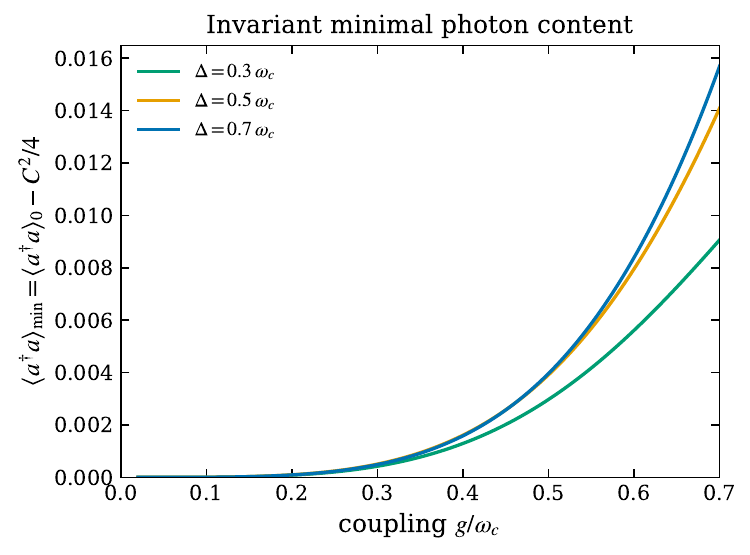}
      \caption{}
      \label{fig:min-photon}
    \end{subfigure}
    \caption{Gauge dependence of the ground-state photon number for
      the $\alpha$-gauge family, with $\omega_{c}=1$ (and $\Delta=0.7$
      where a single value is used). (a) Photon number versus coupling
      $g$ for several values of the gauge parameter $\alpha$: it
      depends strongly on the gauge. (b) Gauge of minimal photon
      number, $\alpha_{\min}=C/2\eta$, versus $g$: it stays strictly
      between the Coulomb ($\alpha=0$) and dipole ($\alpha=1$)
      gauges, starting from the Jaynes-Cummings gauge
      $2\Delta/(\omega_{c}+2\Delta)$, \eqref{eq:weak-coupling}, at
      weak coupling. (c) Photon number versus $\alpha$ for several $g$ (dots
      mark the minima), following the fixed-curvature parabola
      of~\eqref{eq:photon-quadratic}. (d) Minimal photon number
      $\langle\ad a\rangle_{\min}$ versus $g$, reached at
      $\alpha_{\min}=C/2\eta$.}
    \label{fig:photon-gauge}
  \end{figure}

  The corrected Coulomb gauge $\HC=\mathcal H^{(\alpha=0)}$ is reached
  from the standard-gauge Hamiltonian by a composite unitary,
  $|\psi_{\HC}\rangle=\mathcal
  U^{\dagger}U|\psi_{H_{\mathrm{std}}}\rangle$. The first factor,
  $U=\exp[i\frac{\pi}{2}\ad a]$ of~\eqref{eq:rotation-CtoD}, is a
  plain rotation of the field that swaps the quadratures,
  $U^{\dagger}xU=-p$ and $U^{\dagger}pU=x$. The second,
  $\mathcal
  U^{\dagger}=\exp[i\eta\sigma_{x}(a+\ad)]=U_{\lambda}|_{\lambda=\eta}$,
  is a further displacement that depends on the matter and follows the
  same algebra as~\eqref{eq:universal-transform}. Combining these
  two steps gives the corrected-Coulomb QRM gauge expectation values
  directly in terms of those of the standard QRM,
  \begin{align}
    \langle\ad a\rangle_{0} &= \langle\ad a\rangle_{0}^{\mathrm{std}} + \eta\,C^{\mathrm{std}} + \eta^2,
        &
    C &= C^{\mathrm{std}} + 2\eta,
        \label{eq:HC-bridge}
  \end{align}
  where
  $C^{\mathrm{std}}=\langle\sigma_{x}(a+\ad)\rangle_{0}^{\mathrm{std}}$
  is the standard-QRM correlator of~\eqref{eq:C-std}, which can
  therefore also be obtained from the tunable-coherent-state solution.
  The quantities $\langle\ad a\rangle_{0}$ and $C$ on the left are the
  corrected-gauge moments that enter the
  parabola~\eqref{eq:universal-parabola} at $\alpha=0$. Since
  $\Ua=U_{\lambda}|_{\lambda=-\alpha\eta}$
  (Sec.~\ref{subsec:universal-parabola}),
  \eqref{eq:universal-parabola} applies directly with $G=\sigma_{x}$
  and $Q=\langle\sigma_{x}^2\rangle=1$, and gives the photon number
  across the entire gauge family,
  \begin{align}
    \langle\ad a\rangle_{\alpha}
    = \langle\ad a\rangle_{0} - C\eta\,\alpha + \eta^2\alpha^2.
    \label{eq:photon-quadratic}
  \end{align}
  This is the exact photon number of the $\alpha$-gauge QRM, for which we
  note that the coefficient of $\alpha^2$ always equals $\eta^2$. The
  linear coefficient $C$ is the ground-state correlator between the
  matter operator and the field momentum in the corrected Coulomb
  gauge. Figure~\ref{fig:photon-gauge} shows
  how~\eqref{eq:photon-quadratic} depends on $\alpha$, $g$ and
  $\Delta$.
  
  A gauge with no photons would need~\eqref{eq:photon-quadratic}
  to have a real root in $\alpha$, that is, a nonnegative discriminant
  $C^2-4\langle\ad a\rangle_{0}\ge 0$ (because the positive factor
  $\eta^2$ in front of the quadratic term never changes its
  sign). While the discriminant can never be positive
  (Sec.~\ref{subsec:universal-parabola}), numerically we find
  furthermore that it is indeed strictly negative for all couplings
  and detunings that we examined. Over $g\in[0.02,0.7]\,\omega_{c}$
  and $\Delta=0.3,0.5,0.7\,\omega_{c}$ it ranges from about
  $-4\times10^{-8}$ to $-0.063$, and for $\Delta=0.7\,\omega_{c}$ it
  equals $-0.002$, $-0.016$ and $-0.063$ at $g=0.3$, $0.5$ and
  $0.7\,\omega_{c}$. That the discriminant is strictly negative can
  also be shown analytically. By Sec.~\ref{subsec:universal-parabola},
  taking the dipole gauge as reference, a vanishing minimum would
  require $(a+i\lambda\sigma_{x})|\psi\rangle=0$ for the ground state
  $|\psi\rangle$ and some $\lambda$, i.e.,
  $|\psi\rangle=\sum_{s=\pm1}c_{s}|s\rangle|\beta_{s}\rangle$, where
  $|s\rangle$ are the eigenstates of $\sigma_{x}$ and
  $|\beta_{s}\rangle$ are coherent states with $\beta_{s}=-i\lambda s$.
  Inserting this into the Schr\"odinger equation
  for~\eqref{eq:dipole-QRM} and projecting onto $|s\rangle$ gives
  \begin{align}
    (E-\eta^2\omega_{c})\,c_{s}|\beta_{s}\rangle
    =c_{s}\bigl[(\omega_{c}\beta_{s}+isg_{\mathrm{D}})\,\ad|\beta_{s}\rangle
    -isg_{\mathrm{D}}\beta_{s}|\beta_{s}\rangle\bigr]
    +\Delta\,c_{-s}|\beta_{-s}\rangle.
    \label{eq:no-coherent-gs}
  \end{align}
  Since $|\beta\rangle$, $\ad|\beta\rangle$ and $|\beta'\rangle$ are
  linearly independent for $\beta\neq\beta'$ (in the Bargmann
  representation they correspond to $e^{\beta z}$, $z\,e^{\beta z}$
  and $e^{\beta'z}$), this requires $\beta_{s}=-is\eta$ whenever
  $c_{s}\neq0$, so that $\beta_{+1}\neq\beta_{-1}$, and then the
  $\Delta$ term forces $c_{-s}=0$ for both $s$. For $\Delta>0$ such a
  state therefore does not exist. For $\Delta=0$, on the other hand,
  $\HC=\omega_{c}\ad a$ has twofold degenerate ground states with the
  field in its vacuum, so that $\langle\ad a\rangle_{\min}=0$ is
  reached in the Coulomb gauge. We conclude that no gauge
  gives a photon-free ground state, and instead the parabola has a
  unique minimum at $\alpha_{\min} = C/2\eta$, where
  \begin{align}
    \langle\ad a\rangle_{\min} = \langle\ad a\rangle_{0} - \frac{C^2}{4} > 0 .
    \label{eq:min-photon}
  \end{align}
  The value $\langle\ad a\rangle_{\min}$ is the central invariant of
  this analysis, representing a photon content of the ground state
  that no gauge transformation can remove.

  Inserting~\eqref{eq:HC-bridge} into~\eqref{eq:min-photon}, the
  terms with $\eta$ cancel, and with the Hellmann-Feynman relations
  $\langle\ad a\rangle_{0}^{\mathrm{std}}=\partial E_{0}/\partial\omega_{c}$
  and $C^{\mathrm{std}}=\partial E_{0}/\partial g$ for the
  ground-state energy $E_{0}$ of $H_{\mathrm{std}}$ we obtain
  \begin{align}
    \langle\ad a\rangle_{\min}
    =\langle\ad a\rangle_{0}^{\mathrm{std}}-\tfrac14\bigl(C^{\mathrm{std}}\bigr)^2
    =\frac{\partial E_{0}}{\partial\omega_{c}}
    -\frac14\Bigl(\frac{\partial E_{0}}{\partial g}\Bigr)^2,
    \qquad
    \alpha_{\min}=1+\frac{\omega_{c}}{2g}\,\frac{\partial E_{0}}{\partial g}.
    \label{eq:min-spectral}
  \end{align}
  The gauge-invariant minimum is therefore determined by the spectrum
  alone, and can also be obtained from the exact
  solution~\cite{braak_integrability_2011} by implicit
  differentiation. Here $E_{0}$ refers to $H_{\mathrm{std}}$, i.e.,
  without the constant $\eta^2\omega_{c}$, which would otherwise
  contribute to the derivatives. Since $E_{0}$ is a concave function
  of $g$ (as a minimum of expectation values that are linear in $g$)
  and even in $g$, we have $\partial E_{0}/\partial g<0$ for $g>0$, so
  that $\alpha_{\min}<1$ always holds, whereas $\alpha_{\min}>0$ is a
  numerical observation.

  At weak coupling the minimizing gauge can be found analytically. To
  first order in $\eta$, the counter-rotating part of the coupling
  in~\eqref{eq:alpha-def} is proportional to
  $[\alpha\omega_{c}-2(1-\alpha)\Delta](\ad\sigma_{+}-a\sigma_{-})$,
  and second-order perturbation theory gives
  \begin{align}
    \langle\ad a\rangle_{\alpha}
    =\eta^2\,\frac{[\alpha\omega_{c}-2(1-\alpha)\Delta]^2}{(\omega_{c}+2\Delta)^2}+O(\eta^4),
    \qquad
    \alpha_{\min}=\frac{2\Delta}{\omega_{c}+2\Delta}+O(\eta^2).
    \label{eq:weak-coupling}
  \end{align}
  This is the Jaynes-Cummings gauge of Ref.~\cite{stokes_gauge_2019},
  in which the counter-rotating terms vanish to linear order, and
  $\alpha_{\min}$ can be regarded as its nonperturbative
  continuation. The minimum itself is of fourth order. With the
  ground-state energy
  $E_{0}=-\Delta-g^2/(\omega_{c}+2\Delta)-2\Delta g^4/[\omega_{c}(\omega_{c}+2\Delta)^3]+O(g^6)$,
  \eqref{eq:min-spectral} yields
  \begin{align}
    \langle\ad a\rangle_{\min}=\frac{4\Delta^2g^4}{\omega_{c}^2(\omega_{c}+2\Delta)^4}+O(g^6),
    \label{eq:min-weak}
  \end{align}
  in agreement with the discriminant $-4\langle\ad a\rangle_{\min}$
  found numerically at small $g$. At strong coupling, on the other
  hand, $\alpha_{\min}$ decreases toward the Coulomb gauge, and the
  minimum passes through a maximum and decreases again, because the
  tunneling between the two displaced field states of the dipole frame
  is
  suppressed~\cite{twyeffort_irish_generalized_2007,twyeffort_irish_erratum_2007,liu_fundamental_2021}. For
  $\Delta=0.7\,\omega_{c}$ we find $\alpha_{\min}=0.051$ and $0.0016$,
  and $\langle\ad a\rangle_{\min}=0.059$ and $0.014$, at
  $g=1.5\,\omega_{c}$ and $3\,\omega_{c}$, respectively.

  Let us discuss the physical meaning of the minimum photon
  content. In terms of the quadratures $x=a+\ad$ and
  $p=i(\ad-a)$, \eqref{eq:universal-displacement} acts on a
  \emph{single} quadrature. With $\lambda=-\alpha\eta$ and
  $G=\sigma_{x}$ it gives
  \begin{align}
    \Ua^{\dagger}\,x\,\Ua=x,\qquad \Ua^{\dagger}\,p\,\Ua=p-2\alpha\eta\,\sigma_{x} .
    \label{eq:quad-shift}
  \end{align}
  Sweeping the gauge therefore only moves the part of the
  $p$-quadrature that is \emph{correlated with the matter} between
  light and matter. The removable amount
  $C^2/4Q=\langle\sigma_{x}p\rangle^2/4\langle\sigma_{x}^2\rangle$ is
  exactly the field content that a frame moving with the matter can
  absorb, and the gauge of minimal photon number is the one that
  minimizes the variance $\langle p^2\rangle$ of the matter-coupled
  quadrature. No gauge modifies the conjugate quadrature $x$, which
  is gauge invariant by~\eqref{eq:quad-shift}, or the fluctuations
  of $p$ that are not correlated with the matter. The invariant
  minimum is built from these alone,
  \begin{align}
    \langle\ad a\rangle_{\min}=\tfrac14\bigl(\langle x^2\rangle+\langle p^2\rangle_{\min}-2\bigr),
    \qquad \langle p^2\rangle_{\min}=\langle p^2\rangle_{0}-C^2/Q,
  \label{eq:min-quadrature}
  \end{align}
  where the remaining quadrature fluctuations of the field enter once
  every displaceable, matter-coupled contribution has been
  removed. The gauge-invariant quadrature is itself squeezed below the
  vacuum by the counter-rotating interaction at ultrastrong coupling,
  e.g., for the QRM at $g=0.8\,\omega_{c}$, $\Delta=0.7\,\omega_{c}$
  we find $\langle x^2\rangle\simeq0.74<1$.

  This gives the minimum a clear physical meaning and sharpens a
  long-standing question. At ultrastrong coupling the ground state
  contains virtual photons, created together with matter excitations
  by the counter-rotating
  terms. The long debate over whether these photons are ``real'' comes
  from the gauge dependence of their
  number~\cite{de_bernardis_breakdown_2018,di_stefano_resolution_2019,stokes_implications_2022}. We
  can now ask the question more sharply. In each gauge,
  $\langle\ad a\rangle_{\alpha}$ is the expectation value of a
  different field observable, namely, in the corrected Coulomb frame,
  of $(a-\IMI\alpha\eta\sigma_{x})^{\dagger}(a-\IMI\alpha\eta\sigma_{x})$,
  and the minimum belongs to the most vacuum-like of these field
  variables. The photon count in any gauge
  depends on the frame and is partly removable. But it is bounded
  below by a strictly positive number that is the same in every gauge,
  so no ``reference frame'' is free of photons. The removable part is tied to the
  correlator~\eqref{eq:HC-bridge} between matter and field momentum
  and reflects the part of the field that follows the dipole. Note
  that the displacement $U_{\lambda}$ is conditioned on the matter and
  can therefore change the light-matter entanglement, so that the
  minimum is not a measure of entanglement. Rather, the part that
  cannot be removed is the non-coherence of the field conditioned on
  the dipole, i.e., the squeezing and non-Gaussianity of the
  conditional field states, as we show next.

  \subsection{Measurements of the minimum photon number}
  \label{subsec:operational}
  
  It should be possible to measure the minimum photon content, as it
  does not reference any particular gauge, and we now discuss such
  procedures.  Let $G=\sum_{s}s\,\mathcal P_{s}$ be the spectral
  decomposition of the matter coupling operator, with eigenvalues $s$
  and projectors $\mathcal P_{s}$ onto the corresponding matter
  eigenspaces ($s=\pm1$ for the QRM, where $G=\sigma_{x}$). If $G$ is
  measured projectively on the ground state $|\psi_{0}\rangle$ of the
  reference gauge, the outcome $s$ occurs with probability
  $w_{s}=\langle\psi_{0}|\mathcal P_{s}|\psi_{0}\rangle$ and leaves
  the field in the state
  $\rho_{s}=\mathrm{Tr}_{\mathrm{matter}}\bigl[\mathcal
  P_{s}|\psi_{0}\rangle\langle\psi_{0}|\bigr]/w_{s}$. Expectation
  values in $\rho_{s}$ are written as
  $\langle O\rangle_{s}=\mathrm{Tr}[\rho_{s}O]$, the conditional mean
  quadratures are $m^{x}_{s}=\langle x\rangle_{s}$ and
  $m_{s}=\langle p\rangle_{s}$, and the conditional variances are
  $\mathrm{Var}_{s}O=\langle O^2\rangle_{s}-\langle
  O\rangle_{s}^2$. Because $G$ acts only on the matter,
  $U_{\lambda}=\exp[i\lambda G(a+\ad)]$ commutes with every
  $\mathcal P_{s}$, and on the sector $G=s$ it acts on the field as
  the displacement $a\to a+i\lambda s$. It therefore leaves the
  $w_{s}$ and the variances unchanged and only displaces each
  $\rho_{s}$.

  Consider now the following measurement protocol. Measure $G$ (for
  the QRM this is the dipole $\sigma_{x}$) and then measure the field
  quadratures on the post-measurement state. Here we use the centered
  conditional photon number, i.e., for each outcome we take the photon
  number of the conditional state $\rho_{s}$ after subtracting its
  mean field $\langle a\rangle_{s}$, and then average over the
  outcomes,
  \begin{align}
    N_{\mathrm{cond}}=\sum_{s}w_{s}\bigl[\langle\ad a\rangle_{s}-|\langle a\rangle_{s}|^2\bigr]
    =\sum_{s}w_{s}\,\tfrac14\bigl[\mathrm{Var}_{s}x+\mathrm{Var}_{s}p-2\bigr].
    \label{eq:Ncond}
  \end{align}
  Since $U_{\lambda}$ only displaces each $\rho_{s}$, which shifts
  $\langle a\rangle_{s}$ but not the variances, $N_{\mathrm{cond}}$ is
  the same in every gauge, so it is gauge invariant by construction.

  Next the quantity~\eqref{eq:Ncond} can be related to the minimum
  photon content. We use $\ad a=\tfrac14(x^2+p^2-2)$ and split each
  second moment into a variance plus the square of a mean within each
  sector. The shift $p\to p+2\lambda G$ acts as
  $m_{s}\to m_{s}+2\lambda s$, and $x$ does not change, so
  \begin{align}
    \langle\ad a\rangle_{\lambda}=N_{\mathrm{cond}}+\tfrac14\sum_{s}w_{s}\Bigl[(m^{x}_{s})^2+(m_{s}+2\lambda s)^2\Bigr].
  \end{align}
  Minimizing over $\lambda$ gives
  \begin{align}
    \langle\ad a\rangle_{\min}=N_{\mathrm{cond}}+\tfrac14\Bigl[\sum_{s}w_{s}(m^{x}_{s})^2+\min_{\lambda}\sum_{s}w_{s}(m_{s}+2\lambda s)^2\Bigr]\ \ge\ N_{\mathrm{cond}},
    \label{eq:min-vs-Ncond}
  \end{align}
  with equality if and only if the conditional means satisfy
  $m^{x}_{s}=0$ and $m_{s}=-2\lambda_{\ast}s$ for every $s$. In that
  case the conditional field states simply follow the dipole
  classically and differ from one another by a displacement throughout the
  gauge family. The bracket in~\eqref{eq:min-vs-Ncond} is the part
  of the minimum that is not due to the non-coherence of the
  conditional field states.

  For the QRM both conditions hold exactly. The ground state is
  nondegenerate for $\Delta>0$ and invariant under the parity
  $\Pi=\sigma_{z}\exp(i\pi\ad a)$~\cite{fulton_vibronic_1961,braak_symmetries_2019},
  which sends $s\to-s$, $x\to-x$ and
  $p\to-p$. Hence $m^{x}_{-s}=-m^{x}_{s}$, $m_{-s}=-m_{s}$ and
  $w_{+}=w_{-}$. In the dipole frame the antiunitary operator
  $\mathcal T=\sigma_{z}K$ (complex conjugation in the Fock basis)
  also leaves $\mathcal H$ invariant, sends $s\to-s$ and leaves $x$
  unchanged, so $m^{x}_{-s}=m^{x}_{s}$. Together these give
  $m^{x}_{s}=0$. There are only two outcomes, $s=\pm1$, and
  $m_{-1}=-m_{+1}$, so the means automatically have the form
  $m_{s}=-2\lambda_{\ast}s$. For the QRM we therefore have
  \begin{align}
    \langle\ad a\rangle_{\min}=N_{\mathrm{cond}}.
  \end{align}
  Since each conditional field state $\rho_{s}$ is pure,
  $N_{\mathrm{cond}}$ vanishes only if all $\rho_{s}$ are coherent
  states, and the minimum thus measures the non-coherence, i.e., the
  squeezing and non-Gaussianity, of the field conditioned on the
  dipole.

  As a numerical check we computed $N_{\mathrm{cond}}$ directly
  from~\eqref{eq:Ncond} by projecting the exact ground state onto the
  eigenspaces of $G$, and compared it with
  $\langle\ad a\rangle_{\min}=\langle\ad a\rangle_{0}-C^2/4Q$ from the
  parabola. For the QRM at
  $(g,\Delta)/\omega_{c}=(0.3,0.7),(0.6,0.7),(0.8,0.7),(1.0,0.7),(0.8,0.3),(0.5,1.5)$
  they agree to better than $4\times10^{-16}$. For a typical case,
  $g=0.8\,\omega_{c}$ and $\Delta=0.7\,\omega_{c}$, the passive count
  is $0.136$, of which $0.026$ is the invariant minimum and $0.110$ is
  the squared dipole-field correlation term.

  Alternatively, consider a detector that is coupled to the physical
  field, which measures a fixed gauge-covariant operator, one that does
  not change with $\lambda$, as emphasized for Glauber detection in
  Ref.~\cite{PhysRevResearch.3.023079}. In the corrected Coulomb frame
  the vector potential is $A\propto x$ and the transverse electric
  field is $E_{T}=-\dot A$. The Hamiltonian $\mathcal H_{\mathrm{C}}$ depends on
  the field only through $\omega_{c}\ad a$ and through $x$, so
  $\dot x=\IMI[\mathcal H_{\mathrm{C}},x]=\omega_{c}\,p$. Here
  $p=\IMI(\ad-a)$ is the momentum quadrature in this frame, which we
  write $p_{\mathrm{C}}$ to distinguish it from the quadrature in
  other gauges. Hence $E_{T}\propto p_{\mathrm{C}}$. A passive measurement
  of the field noise therefore returns the corrected-Coulomb photon
  number
  $\langle\ad a\rangle_{\alpha=0}=\tfrac14(\langle x^2\rangle+\langle
  p_{\mathrm{C}}^2\rangle-2)$. Equation~\eqref{eq:universal-parabola}
  at $\lambda=0$ then gives the exact decomposition
  \begin{align}
    \langle\ad a\rangle_{\alpha=0}=\langle\ad a\rangle_{\min}+\frac{\langle G\,p_{\mathrm{C}}\rangle^2}{4\langle G^2\rangle},
    \label{eq:passive-decomp}
  \end{align}
  where $\langle G\,p_{\mathrm{C}}\rangle\propto\langle G\,E_{T}\rangle$
  is the equal-time correlator between the dipole and the field. A
  passive field-noise measurement therefore counts more photons than
  the invariant minimum, by exactly the squared dipole-field
  correlation. This correlation can itself be measured by
  cross-correlating the field noise with the dipole. The invariant
  minimum is the passive count with the dipole-following contribution
  removed. Note that this needs a projective measurement of the
  matter dipole, or equivalently the joint field-dipole statistics.
  Here we assumed that the detector responds to the transverse
  electric field. In the multipolar description, however, the total
  field outside the matter is given by the displacement field
  $D_{T}/\varepsilon_{0}$, which corresponds to the dipole-gauge
  momentum quadrature, and such a detector would instead record
  $\langle\ad a\rangle_{\alpha=1}$, e.g., $0.246$ instead of $0.136$
  for $g=0.8\,\omega_{c}$ and $\Delta=0.7\,\omega_{c}$. By contrast,
  $N_{\mathrm{cond}}$ does not depend on this choice, because the two
  fields differ by a shift proportional to the dipole. We also note
  that a photodetector of Glauber type does not register the virtual
  photons of the ground state at all. They can, however, be converted
  into real photons with the help of additional matter
  levels~\cite{stassi_spontaneous_2013,giannelliIntegratedConversionPhotodetection2023,PhysRevResearch.6.013008},
  or probed spectroscopically with an ancillary
  qubit~\cite{lolli_ancillary_2015}. Similarly, the conditional
  protocol requires measurements on time scales short compared to
  $1/\omega_{c}$, $1/\Delta$ and $1/g$, since otherwise the result
  depends on how the measurement is switched
  on~\cite{PhysRevResearch.3.023079}.

  \section{Dicke model and role of the dipole self-energy}
  \label{sec:dicke}
  
  \subsection{Gauge family and curvature}
  \label{subsec:dicke-gauges}

  The natural multi-emitter setting is the Dicke
  model~\cite{dicke_coherence_1954,kirton_introduction_2019}, in which
  $N_{s}$ two-level systems couple collectively to one mode, and to
  which also many correlated light-matter systems without
  superradiance can be mapped~\cite{schellenberger_almost_2024}. Its
  gauge-invariant form follows from the same minimal-coupling
  procedure as for the QRM~\cite{garziano_gauge_2020}. The
  Coulomb-gauge Hamiltonian is generated from the bare matter
  Hamiltonian by the collective transformation
  $\hat{\mathcal{U}}_{N}=\exp[2i\eta(a+\ad)\hat J_{x}]$, where
  $\hat J_{\alpha}=\tfrac12\sum_{l}\sigma^{\alpha}_{l}$ is the
  collective spin and $\eta$ is the single-emitter coupling. This
  gives the form, valid to all orders in the field,
  \begin{align}
    \hat{\mathcal H}_{\mathrm{cg}} = \omega_{c}\,\ad a
    + \omega_{x}\bigl\{\hat J_{z}\cos[2\eta(a+\ad)] + \hat J_{y}\sin[2\eta(a+\ad)]\bigr\},
    \label{eq:dicke-coulomb}
  \end{align}
  The dipole gauge,
  $\hat{\mathcal H}_{\mathrm{dg}}=\hat{\mathcal
    U}_{N}^{\dagger}\hat{\mathcal H}_{\mathrm{cg}}\hat{\mathcal
    U}_{N}$, is the bilinear Hamiltonian
  \begin{align}
  \hat{\mathcal H}_{\mathrm{dg}} = \omega_{c}\,\ad a + \omega_{x}\hat J_{z}
  + 2i\eta\omega_{c}(\ad-a)\hat J_{x} + 4\eta^2\omega_{c}\hat J_{x}^2.
  \label{eq:dicke-dipole}
  \end{align}
  The intermediate $\alpha$-gauge then results from applying only a
  fraction $\alpha$ of the transformation,
  $\hat{\mathcal{U}}_{\alpha}=\exp[2i\alpha\eta(a+\ad)\hat J_{x}]$, with
  $\hat{\mathcal H}^{(\alpha)}=\hat{\mathcal
    U}_{\alpha}^{\dagger}\,\hat{\mathcal
    H}_{\mathrm{cg}}\,\hat{\mathcal U}_{\alpha}$ and
  $\hat{\mathcal U}_{\alpha}^{\dagger}\,a\,\hat{\mathcal
    U}_{\alpha}=a+2i\alpha\eta\hat J_{x}$. It interpolates between the
  Coulomb ($\alpha=0$) and dipole ($\alpha=1$) gauges,
  \begin{align}
    \hat{\mathcal H}^{(\alpha)} =\;& \omega_{c}\,\ad a
                                     + 2i\alpha\eta\omega_{c}(\ad-a)\hat J_{x} + 4\alpha^2\eta^2\omega_{c}\hat J_{x}^2 \nonumber\\
                                   &+ \omega_{x}\bigl\{\hat J_{z}\cos[2(1-\alpha)\eta(a+\ad)] + \hat J_{y}\sin[2(1-\alpha)\eta(a+\ad)]\bigr\}.
                                     \label{eq:dicke-alpha-gauge}
  \end{align}
  The linear coupling and the dipole self-energy then appear as the
  first- and second-order terms of the displacement,
  \begin{align}
    \hat{\mathcal U}_{\alpha}^{\dagger}\,\ad a\,\hat{\mathcal U}_{\alpha}
    = \ad a + 2i\alpha\eta\,\hat J_{x}(\ad-a) + 4\alpha^2\eta^2\hat J_{x}^2,
    \label{eq:dicke-disp}
  \end{align}
  so the self-polarization $4\alpha^2\eta^2\omega_{c}\hat J_{x}^2$
  equals the squared part of the displacement. The maps close a
  commuting square,
  $\hat{\mathcal H}_{\mathrm{cg}}=\hat{\mathcal U}_{N}\hat{\mathcal
    H}_{\mathrm{dg}}\hat{\mathcal U}_{N}^{\dagger}$ with
  $\hat{\mathcal H}^{(0)}=\hat{\mathcal H}_{\mathrm{cg}}$ and
  $\hat{\mathcal H}^{(1)}=\hat{\mathcal H}_{\mathrm{dg}}$, and because
  each $\hat{\mathcal U}_{\alpha}$ is unitary, every member of the
  corrected family has the same spectrum.

  The transformation $\hat{\mathcal U}_{\alpha}$ is exactly the
  displacement of~\eqref{eq:universal-U}, with
  $\lambda=-\alpha\eta\sqrt{N_{s}}=-\alpha g$, where
  $g=\eta\sqrt{N_{s}}$ is the collective coupling, and the collective
  matter operator
  $G=N_{s}^{-1/2}\sum_{l}\sigma^{x}_{l}=2N_{s}^{-1/2}\hat J_{x}$
  (Sec.~\ref{subsec:universal-parabola}). The minimum photon content
  parabola thus also occurs for the Dicke model, with curvature
  \begin{align}
    Q = \langle G^2\rangle = 1 + \frac{1}{N_{s}}\sum_{l\neq l'}\langle\sigma^{x}_{l}\sigma^{x}_{l'}\rangle.
  \label{eq:dicke-Q}
  \end{align}
  For $N_{s}=1$ we have $Q=1$ exactly, because $G^2$ is the identity
  and the self-polarization term is just a constant. For $N_{s}>1$ the
  curvature measures the collective dipole fluctuation.

  \subsection{Standard Dicke model, self-energy, and the no-go theorem}
  \label{subsec:dicke-standard}

  The magnitude of this fluctuation, and whether the gauges are
  unitarily equivalent at all, depend on the dipole self-energy that
  the displacement produces, as we now discuss. Consider the
  \emph{standard} Dicke model, with bilinear coupling and without the
  dipole self-energy,
  \begin{align}
    \hat{\mathcal H}_{\mathrm{std}}
    = \omega_{c}\ad a + \omega_{x}\hat J_{z}
    + 2i\eta\omega_{c}(\ad-a)\hat J_{x},
    \label{eq:dicke-uncorrected}
  \end{align}
  i.e., \eqref{eq:dicke-dipole} without the term
  $4\eta^2\omega_{c}\hat J_{x}^2$. It is related to the form in widest
  use, with coupling $(a+\ad)\hat J_{x}$, by the
  rotation~\eqref{eq:rotation-CtoD}. In the Coulomb gauge the
  self-energy corresponds to the diamagnetic term: expanding the
  cosine in~\eqref{eq:dicke-coulomb} to second order in $\eta$ gives
  $-2\eta^2\omega_{x}(a+\ad)^2\hat J_{z}$, which for
  $\hat J_{z}\to-N_{s}/2$ becomes $D(a+\ad)^2$ with the
  Thomas-Reiche-Kuhn value
  $D=N_{s}\eta^2\omega_{x}$~\cite{savasta_thomasreichekuhn_2021}.
  Beyond this order, and at finite $N_{s}$, the trigonometric
  structure of~\eqref{eq:dicke-coulomb} is needed, and no constant $D$
  makes a bilinear Coulomb-gauge model a member of the corrected
  family. Without the self-energy,
  $\hat{\mathcal H}_{\mathrm{std}}$ is \emph{not} the
  image
  $\hat{\mathcal U}_{\alpha}^{\dagger}\hat{\mathcal
    H}_{\mathrm{cg}}\hat{\mathcal U}_{\alpha}$ of the corrected family
  for any $\alpha$. No gauge is therefore unitarily equivalent to it,
  and its spectrum is different~\cite{garziano_gauge_2020}. As a check
  of the numerics we verified that the corrected Coulomb and dipole
  Hamiltonians, which are unitarily equivalent by construction, have
  the same spectra and give the same $\langle\hat J_{x}^2\rangle$.

  For the QRM, i.e., $N_{s}=1$, on the other hand, the collective
  dipole is $\hat J_{x}=\sigma_{x}/2$, so $\hat J_{x}^2=\tfrac14$ is a
  number and the self-energy term
  $4\alpha^2\eta^2\omega_{c}\hat J_{x}^2=\alpha^2\eta^2\omega_{c}$
  reduces to a gauge-dependent constant. It shifts every gauge by the
  same amount and drops out of every commutator. The unitary square
  then closes automatically, and the corrected-Coulomb and dipole Rabi
  Hamiltonians are exactly unitarily equivalent whether or not the
  constant is kept. Nevertheless the bilinear Coulomb-gauge QRM is not
  gauge consistent, because it lacks the trigonometric structure
  of~\eqref{eq:corrected-Coulomb}, so that the gauge problem is not
  only a matter of the self-energy.
  
  The same operator that restores gauge invariance also enforces the
  superradiant no-go
  theorem~\cite{hepp_superradiant_1973,rzazewski_phase_1975,nataf_nogo_2010}. With
  the self-energy, the dipole-gauge
  Hamiltonian~\eqref{eq:dicke-dipole} can be written as
  $\omega_{c}(a+2i\eta\hat J_{x})^{\dagger}(a+2i\eta\hat
  J_{x})+\omega_{x}\hat J_{z}$, so that its ground-state energy can
  never fall below the value $-\omega_{x}N_{s}/2$ without
  coupling. The energy that a static dipole would gain from the
  light-matter coupling is thus exactly compensated, and no
  symmetry-breaking macroscopic photon occupation can
  form~\cite{jaako_ultrastrong-coupling_2016,garziano_gauge_2020}. In
  the Coulomb gauge the same role is played by the diamagnetic term at
  its TRK value. Without the self-energy the compensation is lost, and
  the standard Dicke model does show a superradiant transition. This
  transition is an artifact of dropping the self-energy (or the
  diamagnetic term in the Coulomb gauge) and not a physical
  instability, at least for a spatially uniform single-mode field with
  minimal coupling~\cite{andolina_cavity_2019}; for circuit QED and
  for spatially varying fields the question has been
  debated~\cite{nataf_nogo_2010,viehmann_superradiant_2011,andolina_theory_2020}. The
  self-energy $4\eta^2\omega_{c}\hat J_{x}^2$ thus cures two
  problems. It makes the gauges unitarily equivalent \emph{and}
  forbids superradiance. This stabilization shows up in the gauge
  parabola, the curvature $Q=\langle G^2\rangle$ of which grows with
  coupling without the term and stays near one with it, as we show
  next.

  \subsection{Curvature from exact diagonalization}
  \label{subsec:dicke-curvature}

\begin{figure}[t]
    \centering
    \begin{subfigure}{0.32\textwidth}
      \centering
      \includegraphics[width=\linewidth]{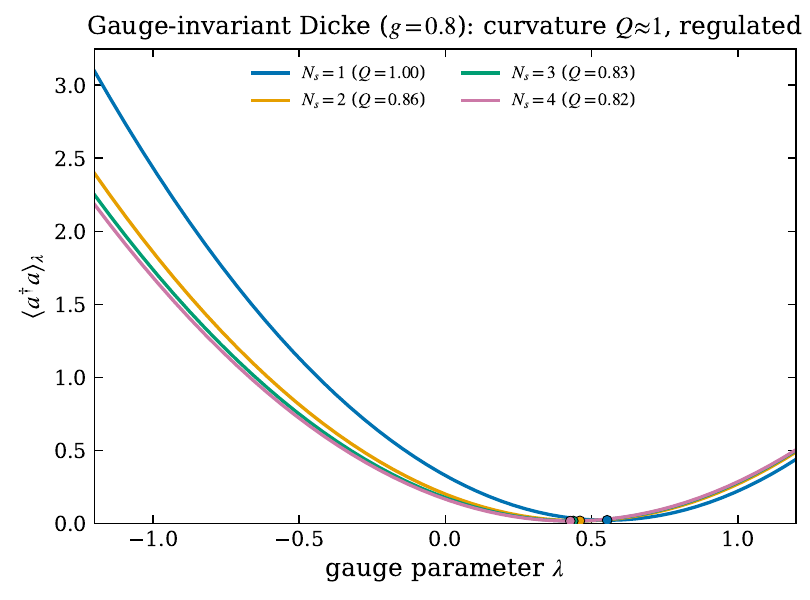}
      \caption{}
      \label{fig:dicke-parabolas}
    \end{subfigure}\hfill
    \begin{subfigure}{0.32\textwidth}
      \centering
      \includegraphics[width=\linewidth]{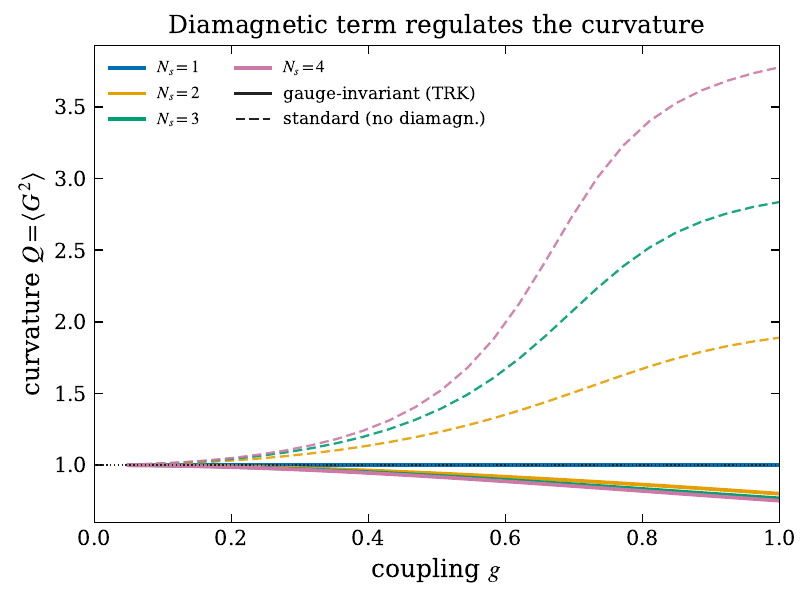}
      \caption{}
      \label{fig:dicke-curvature}
    \end{subfigure}\hfill
    \begin{subfigure}{0.32\textwidth}
      \centering
      \includegraphics[width=\linewidth]{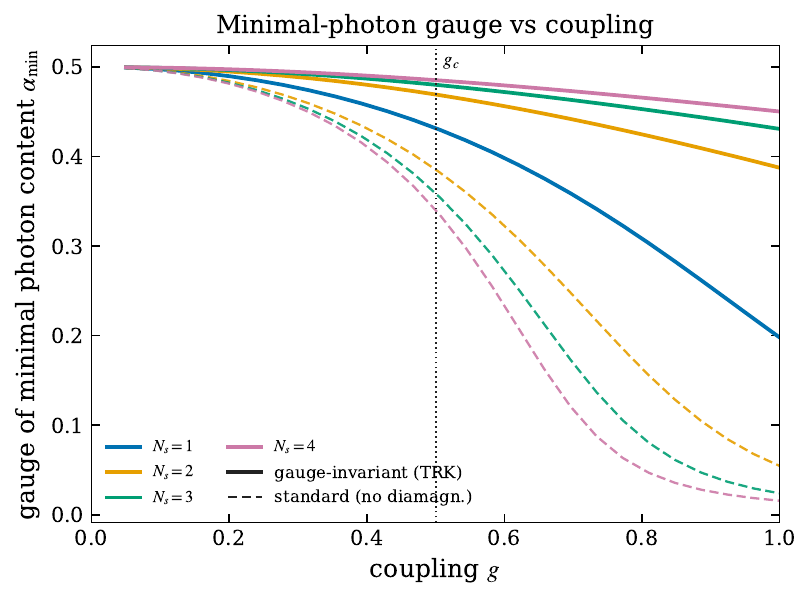}
      \caption{}
      \label{fig:dicke-alpha-min}
    \end{subfigure}

  \vspace{0.6em}

  \begin{subfigure}{0.32\textwidth}
    \centering
    \includegraphics[width=\linewidth]{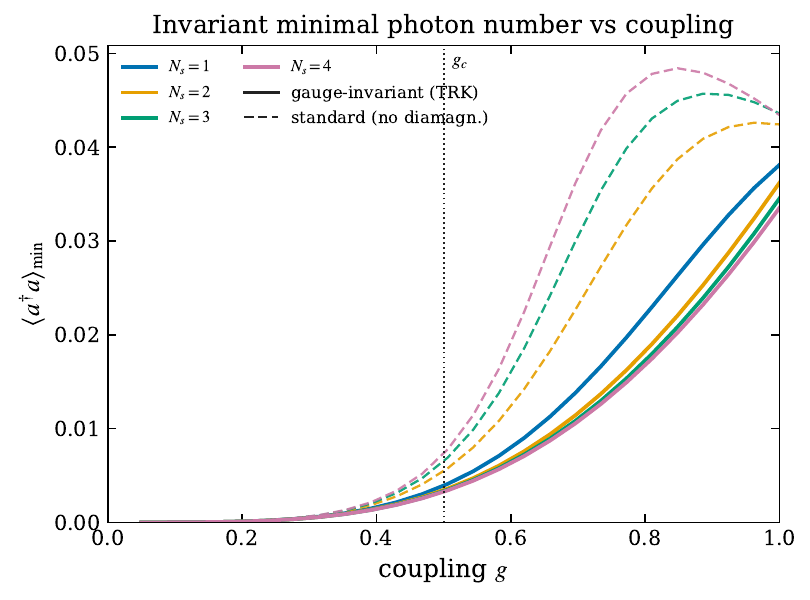}
    \caption{}
    \label{fig:dicke-min-photon}
  \end{subfigure}\hfill
  \begin{subfigure}{0.64\textwidth}
    \centering
    \includegraphics[width=\linewidth]{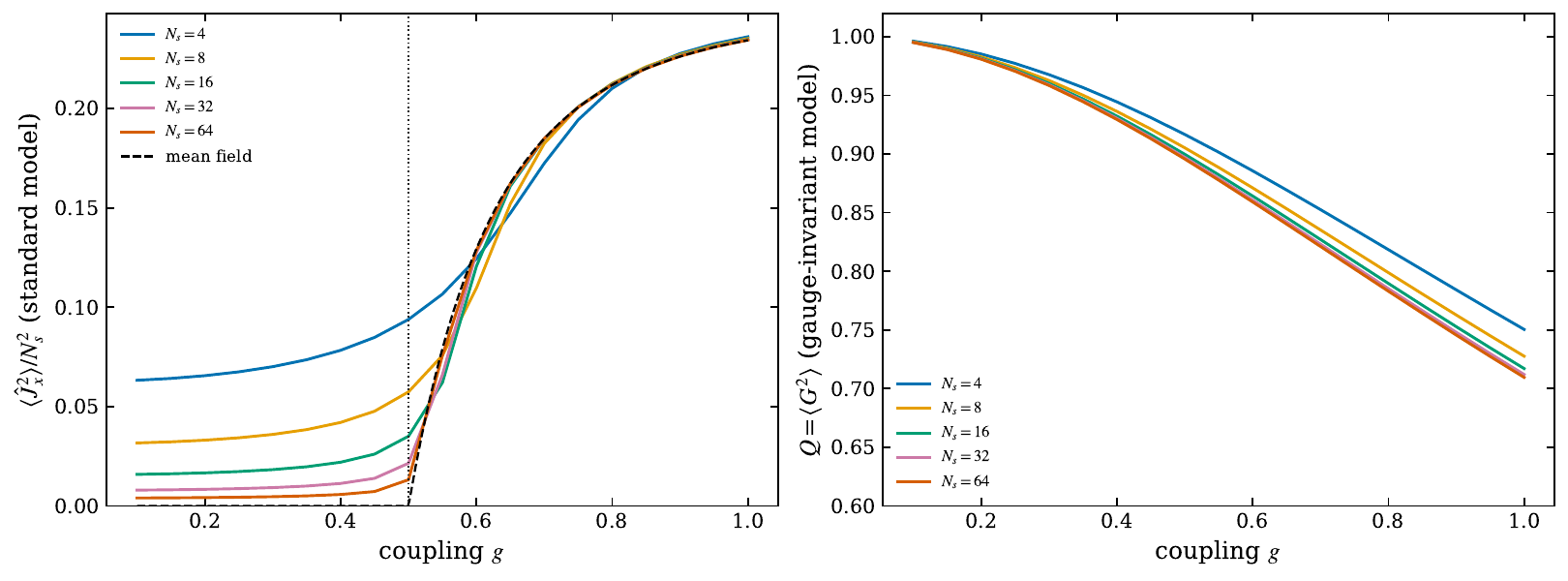}
    \caption{}
    \label{fig:dicke-spinj}
  \end{subfigure}
  \caption{The Dicke model across gauges (exact diagonalization,
    $\omega_{c}=\omega_{x}=1$). (a) Photon number across gauges for
    $N_{s}=1$-$4$ at $g=0.8\,\omega_{c}$ in the gauge-invariant model;
    dots mark the minima. The photon number is an exact parabola in
    the gauge parameter for every $N_{s}$,
    and~\eqref{eq:universal-parabola} matches the numerics to machine
    precision. (b) Curvature $Q=\langle G^2\rangle$ versus coupling,
    for the gauge-invariant model (solid) and the standard model
    (dashed). $Q$ is an ordinary ground-state moment and is well
    defined for either Hamiltonian. It equals one for $N_{s}=1$. For
    $N_{s}>1$ it grows with coupling in the standard model, while the
    dipole self-energy keeps it near one, slightly below. This is the
    same term that forbids superradiance through the no-go
    theorem. (c) Gauge of minimal photon number,
    $\alpha_{\min}=1+C_{\mathrm{D}}/(2Qg)$ ($\alpha=0$ Coulomb,
    $\alpha=1$ dipole), where $C_{\mathrm{D}}=\langle G\,p\rangle$ is
    the correlator in the dipole-gauge ground state. For the
    gauge-invariant model (solid) it is a true invariant of the gauge
    orbit; in the thermodynamic limit it equals $1/2$ for all $g$,
    see~\eqref{eq:dicke-thermo}. The dashed curves apply the same
    formula to the standard model and are a formal quantity only,
    since that model has no gauge orbit. (d) Minimal photon number
    $\langle\ad a\rangle_{\min}=\langle\ad a\rangle_{0}-C^2/4Q$ for
    the gauge-invariant model (solid). The dashed curves are the same
    expression applied to the standard model and are not gauge
    invariant. In (c) and (d) the dotted vertical line marks the
    mean-field superradiant critical coupling of the standard model,
    $g_{c}=\tfrac12\sqrt{\omega_{x}/\omega_{c}}=0.5$, and the dashed
    curves turn over nonmonotonically near it, which reflects the
    superradiant instability of that model. (e) Scaling to large
    $N_{s}$ in the permutation-symmetric sector $j=N_{s}/2$, for
    $N_{s}=4$-$64$. Left: for the standard model,
    $\langle\hat J_{x}^2\rangle/N_{s}^2$ falls onto the mean-field
    curve $\tfrac14[1-(g_{c}/g)^{4}]$ (dashed) above $g_{c}$ and
    decays as $1/N_{s}$ below it. Right: for the gauge-invariant model
    the curvature $Q$ converges to the finite limit $Q_{\infty}$
    of~\eqref{eq:dicke-thermo} at every coupling.\label{fig:dicke}}
  \end{figure}

  We test this by exact diagonalization (for exact solutions of the
  Dicke model at finite $N_{s}$ see Ref.~\cite{he_exact_2015}). The bosonic space is
  infinite-dimensional, so ED needs a finite photon cutoff. We use an
  adaptive cutoff
  $N_{\mathrm{cutoff}}(g)=\max\bigl(24,\lfloor20+34\,g/\omega_{c}\rfloor\bigr)$,
  which runs from $24$ at the weakest coupling shown to $54$ at
  $g=\omega_{c}$, and we check convergence directly: for $N_{s}=4$,
  adding $14$ more Fock states changes $\langle\ad a\rangle_{0}$, $C$
  and $Q$ by at most a few times $10^{-15}$, which is machine
  precision. Building the gauge family with the displacement
  $\hat{\mathcal U}_{\alpha}$ and evaluating $\langle\ad a\rangle$
  reproduces the parabola~\eqref{eq:universal-parabola} to machine
  precision for every $N_{s}$ (Fig.~\ref{fig:dicke-parabolas}), as it
  must, which serves as a check of the numerics. The
  curvature, however, depends on whether the self-energy is kept,
  and this is where the two models part
  (Fig.~\ref{fig:dicke-curvature}). In the \emph{standard} Dicke
  model, with bilinear coupling and no self-energy, nothing
  restrains the collective dipole fluctuation. The curvature grows
  with coupling and emitter number, $Q\simeq1.89,\,2.84,\,3.78$ for
  $N_{s}=2,3,4$ at $g=\omega_{c}$, following the superradiant tendency
  of this model (which is not gauge invariant). Restoring the
  self-energy, which makes the gauges unitarily
  mappable, changes this. The term that forbids the superradiant
  transition also bounds the fluctuation, so for the gauge-invariant
  model $Q$ stays close to one and drops slightly below it
  ($Q\simeq0.80,0.77,0.75$ for $N_{s}=2,3,4$ at $g=\omega_{c}$). The
  curvature of the gauge parabola is therefore a sensitive probe of
  the self-energy: it grows when the term is absent and stays
  regulated when it is present, which separates the standard model
  from its gauge-invariant version.

  \subsection{Minimal photon number and minimizing gauge}
  \label{subsec:dicke-minimum}

  The analysis we used for the QRM carries over directly to the actual
  gauge family of the Dicke model. Figures~\ref{fig:dicke-alpha-min}
  and~\ref{fig:dicke-min-photon} (solid curves) show, as a function of
  coupling, the gauge at which the minimum is reached and the minimum
  itself. As for the QRM before, the minimum is strictly positive. In
  each sector of fixed total spin, a vanishing minimum would require a
  ground state $\sum_{m}c_{m}|m\rangle|\beta_{m}\rangle$, with
  eigenstates $|m\rangle$ of $\hat J_{x}$ and coherent states
  $|\beta_{m}\rangle$. The field terms of~\eqref{eq:dicke-dipole} then
  require $\beta_{m}=-2i\eta m$, which differ for different $m$, and
  since $\omega_{x}\hat J_{z}$ connects $m$ with $m\pm1$, all $c_{m}$
  must vanish for $\omega_{x}>0$. Since $\lambda=-\alpha g$ for the Dicke family
  (Sec.~\ref{subsec:universal-parabola}), the minimizing gauge
  fraction is
  $\alpha_{\min}=-\lambda_{\ast}/g=1+C_{\mathrm{D}}/(2Qg)$, where
  $C_{\mathrm{D}}=\langle G\,p\rangle$ is evaluated in the
  dipole-gauge ground state. Equivalently, $\alpha_{\min}=C/(2Qg)$
  with the corrected-Coulomb correlator $C=C_{\mathrm{D}}+2Qg$. This
  is the Dicke counterpart of $\alpha_{\min}=C/(2\eta)$ for the QRM
  (Sec.~\ref{subsec:photon-gauge}).  The gauge-invariant Dicke model
  belongs to a family of unitarily equivalent gauges, whereas the
  standard model has no such family and does not correspond to
  any $\alpha$ of the corrected gauge family. The dashed curves in
  Figs.~\ref{fig:dicke-alpha-min} and~\ref{fig:dicke-min-photon} are
  therefore not an invariant minimum. They only show what the same
  algebraic extremum gives when we apply the formula formally to the
  ground state of the standard model, with the same collective
  operator $G$ that generates the corrected gauge family. Both panels
  mark the mean-field superradiant critical coupling of the standard
  model, $g_{c}=\tfrac12\sqrt{\omega_{x}/\omega_{c}}=0.5$. We obtain
  it from a Holstein-Primakoff
  linearization~\cite{holstein_field_1940} of the
  Hamiltonian~\eqref{eq:dicke-uncorrected}, followed by a Bogoliubov
  stability analysis. At the finite $N_{s}$ studied here this is a
  crossover and not a sharp transition, but it leaves a visible mark
  in both formal quantities. The formal $\alpha_{\min}$ of the
  standard model starts from the same equal-partition value $1/2$ at
  weak coupling as the genuine one, but falls much faster with $g$,
  toward the Coulomb gauge (Fig.~\ref{fig:dicke-alpha-min}), following
  the unregulated growth of the curvature in
  Fig.~\ref{fig:dicke-curvature}. The difference is clearer for the
  minimum itself (Fig.~\ref{fig:dicke-min-photon}). For the
  gauge-invariant model the true invariant minimum increases smoothly
  with coupling for every $N_{s}$ in the range shown. At stronger
  coupling it also passes through a maximum, e.g., at
  $g\approx1.2$, $1.7$ and $2.6\,\omega_{c}$ for $N_{s}=1$, $2$ and
  $4$, but this maximum moves to larger coupling as $N_{s}$ grows and
  is absent in the thermodynamic limit, see~\eqref{eq:dicke-thermo}.
  For the standard model the formal
  analogue rises and then turns over at a
  coupling that instead moves closer to $g_{c}$ as $N_{s}$ grows. This is a
  finite-size sign of the superradiant instability that only the
  standard model has, even though the plotted quantity has no
  gauge-invariant meaning for that model. The gauge-invariant model
  shows nothing like this: the same self-energy that regulates the
  curvature and forbids superradiance also keeps its true invariant
  minimum smooth.

  The measurement scheme of Sec.~\ref{subsec:operational} carries over
  to the Dicke model with one difference. The symmetries still give
  $m^{x}_{s}=0$ and $m_{-s}=-m_{s}$, but $G$ now has many eigenvalues
  and $m_{s}$ need not be linear in $s$. In general we therefore only
  have $\langle\ad a\rangle_{\min}\ge N_{\mathrm{cond}}$, with
  equality when $m_{s}$ is linear. We checked numerically that the two
  agree to $6\times10^{-16}$ at $N_{s}=2$, where the three-valued
  spectrum of $G$ and parity force $m_{s}$ to be linear, and to
  $8\times10^{-7}$ at $N_{s}=3$, $g=0.8\,\omega_{c}$. The small
  positive difference is the nonlinearity of $m_{s}$
  that~\eqref{eq:min-vs-Ncond} allows.

  \subsection{Collective polarization and the order parameter}
  \label{subsec:dicke-order}
  
  The observations of Sec.~\ref{subsec:dicke-minimum} read the
  approaching instability from a quantity derived from the photon
  number and not from the matter order parameter itself. The natural
  order parameter of the superradiant transition is the collective
  polarization $\langle\hat J_{x}\rangle$, which is macroscopic in the
  superradiant phase and zero in the normal phase. At any finite
  $N_{s}$, however, this mean vanishes \emph{identically}, for both
  the standard and the gauge-invariant Hamiltonian. Both commute with
  the $\mathbb{Z}_{2}$ parity
  $\hat\Pi=\exp[i\pi(\ad a+\hat J_{z}+N_{s}/2)]$, under which
  $\hat a\to-\hat a$ and
  $\hat J_{x},\hat J_{y}\to-\hat J_{x},-\hat J_{y}$. Every term of
  Eqs.~\eqref{eq:dicke-uncorrected} and~\eqref{eq:dicke-coulomb} is
  invariant under this operation, while $\hat J_{x}$ itself is
  odd. The nondegenerate ground state is therefore a parity eigenstate
  with $\langle\hat J_{x}\rangle=0$ exactly, which we confirm
  numerically to $10^{-20}$ at every coupling and $N_{s}$ we examined,
  for both models. Spontaneous symmetry breaking, and a nonzero
  $\langle\hat J_{x}\rangle$ with it, appears only for
  $N_{s}\to\infty$. Instead at finite $N_{s}$ its approach shows up in
  the fluctuation $\langle\hat J_{x}^2\rangle$. The intensive density
  $\langle\hat J_{x}^2\rangle/N_{s}^2$ is tied to the curvature by
  $Q=\langle G^2\rangle=4\langle\hat J_{x}^2\rangle/N_{s}$ (with
  \eqref{eq:dicke-Q}, since $G=2\hat J_{x}/\sqrt{N_{s}}$). We directly
  evaluated this density numerically. For the standard model at
  $N_{s}=4$ it rises from $0.062$ at weak coupling to $0.094$ at
  $g=g_{c}$ and $0.236$ at $g=\omega_{c}$, and at $g=\omega_{c}$ it
  has the same value, $0.236$, for $N_{s}=2,3,4$. This is the
  finite-size sign of the growing collective polarization fluctuation
  that would condense into a macroscopic $\langle\hat J_{x}\rangle$ in
  the thermodynamic limit. For the gauge-invariant model it instead
  falls over the same range, from $0.062$ to $0.057$ and $0.047$ at
  $N_{s}=4$, and from $0.125$ to $0.118$ and $0.100$ at
  $N_{s}=2$. This confirms directly, in the matter sector and not
  through the photon number, that only the standard model develops the
  instability that the no-go theorem forbids.

  \subsection{Large $N_{s}$ and thermodynamic limit}
  \label{subsec:dicke-thermo}

  Since $N_{s}\le4$ cannot by itself establish a thermodynamic
  statement, we repeat the calculation in the permutation-symmetric
  sector $j=N_{s}/2$. This sector contains the ground state and has
  dimension $N_{s}+1$ instead of $2^{N_{s}}$, so we can reach
  $N_{s}=64$ at fixed collective coupling $g$
  (Fig.~\ref{fig:dicke}(e)). For $N_{s}=2$-$4$ it reproduces the
  full-basis $Q$ to $10^{-15}$ for both models, and we raise the
  photon cutoff with $N_{s}$ and $g$ until $Q$ is stable to better
  than $10^{-9}$. For the standard model,
  $\langle\hat J_{x}^2\rangle/N_{s}^2$ vanishes as $1/N_{s}$ for
  $g<g_{c}$, grows slowly at $g_{c}$, and for $g>g_{c}$ converges to a
  finite value, so $Q$ grows linearly with $N_{s}$ (at $g=\omega_{c}$,
  $Q/N_{s}=0.9392,0.9383,0.9379$ for $N_{s}=16,32,64$, approaching the
  mean-field value $0.9375$). The limit agrees with the mean-field
  order parameter. Minimizing the classical energy
  $E/N_{s}=-\omega_{c}g^2\sin^2\theta-\tfrac{\omega_{x}}{2}\cos\theta$
  over the spin angle gives
  $\langle\hat
  J_{x}\rangle^2/N_{s}^2=\tfrac14\bigl[1-(g_{c}/g)^{4}\bigr]$ with
  $g_{c}=\tfrac12\sqrt{\omega_{x}/\omega_{c}}$. At $g=\omega_{c}$ we
  find $0.2345$ against $0.2344$, and at $g=0.8\,\omega_{c}$ we find
  $0.2120$ against $0.2119$. For the gauge-invariant model the same
  density decays as $1/N_{s}$ at every coupling, and $Q$ converges to
  a finite limit (at $g=0.8\,\omega_{c}$,
  $Q=0.818,0.799,0.790,0.785,0.783$ for $N_{s}=4,8,16,32,64$). There
  is no condensation at any $g$. The no-go theorem is thus confirmed
  at system sizes where mean-field theory already reproduces the
  transition of the standard model quantitatively, while the
  gauge-invariant curvature stays $O(1)$.

  These numbers can be compared with the exact thermodynamic limit of
  the gauge-invariant model. With the Holstein-Primakoff
  representation $\hat J_{x}\simeq\tfrac12\sqrt{N_{s}}\,X_{b}$ and
  $\hat J_{z}\simeq-N_{s}/2+b^{\dagger}b$~\cite{holstein_field_1940},
  where $X_{b}=b+b^{\dagger}$ and $P_{b}=\IMI(b^{\dagger}-b)$, the
  dipole-gauge Hamiltonian~\eqref{eq:dicke-dipole} becomes
  \begin{align}
    \hat{\mathcal H}_{\mathrm{dg}}\simeq\frac{\omega_{c}}{4}\bigl[x^2+(p+2gX_{b})^2\bigr]
    +\frac{\omega_{x}}{4}\bigl(X_{b}^2+P_{b}^2\bigr)+\text{const},
  \end{align}
  which is quadratic and positive definite for all $g$. Its Gaussian
  ground state yields, with $\Sigma=\omega_{c}+\omega_{x}$ and
  $R=\sqrt{\Sigma^2+4\omega_{c}\omega_{x}g^2}$,
  \begin{align}
    Q_{\infty}=\frac{\Sigma}{R},\qquad
    \alpha_{\min}=\frac{\omega_{x}}{\omega_{c}+\omega_{x}},\qquad
    \langle\ad a\rangle_{\min}=\frac{(R-\Sigma)^2}{4R\Sigma},
    \label{eq:dicke-thermo}
  \end{align}
  e.g., $Q_{\infty}=0.781$ at $g=0.8\,\omega_{c}$ for
  $\omega_{x}=\omega_{c}$, which the values above approach as
  $1/N_{s}$. Remarkably, in the thermodynamic limit the gauge of
  minimal photon number is the Jaynes-Cummings gauge
  $\omega_{x}/(\omega_{c}+\omega_{x})$ for all couplings, so that the
  coupling dependence of $\alpha_{\min}$ in
  Fig.~\ref{fig:dicke-alpha-min} is a finite-size effect. The
  minimum~\eqref{eq:dicke-thermo} increases monotonically with $g$,
  and at weak coupling it reduces to
  $g^4\omega_{c}^2\omega_{x}^2/\Sigma^4$, the same
  as~\eqref{eq:min-weak} for the QRM with $2\Delta\to\omega_{x}$ and
  $\eta\to g$.

  \section{Conclusion and outlook}
  \label{sec:conclusion}

  We have shown that the gauge dependence of the ground-state photon
  number in single-mode light-matter models obeys an exact law.
  Whenever the gauge change displaces the field by an amount that is
  proportional to a matter operator $G$, the photon number is a
  parabola in the gauge parameter, with the fluctuation
  $Q=\langle G^2\rangle$ entering as curvature and the correlator
  $C=\langle G\,p\rangle$ as linear coefficient. Its minimum,
  $\langle\ad a\rangle_{\min}=\langle\ad a\rangle_{0}-C^2/4Q$, is the
  same in every gauge and is that part of the photon number which has
  gauge-independent meaning. For the QRM it is determined by the
  ground-state energy~\eqref{eq:min-spectral} alone, so that it
  follows from the exact
  solution~\cite{braak_integrability_2011,xieQuantumRabiModel2017},
  which we used here in the form of tunable coherent
  states~\cite{chen_exact_2011,xie_exact_2019}. We proved that the
  minimum is strictly positive for the Rabi and Dicke models with
  level splitting, since no gauge brings the ground state into a
  product state with the field in its vacuum. The gauge of minimal
  photon number lies in between the Coulomb and the dipole gauge. At
  weak coupling, it is precisely the Jaynes-Cummings gauge of
  Ref.~\cite{stokes_gauge_2019}, while at very strong coupling it
  approaches the Coulomb gauge, as the tunneling between the displaced
  oscillator states of the dipole
  frame~\cite{twyeffort_irish_generalized_2007,twyeffort_irish_erratum_2007}
  is suppressed.

  For the Dicke model the curvature is the collective dipole
  fluctuation, which is controlled by the dipole self-energy. The
  presence of the latter term makes the gauges unitarily equivalent
  and enforces the no-go theorem for
  superradiance~\cite{hepp_superradiant_1973,garziano_gauge_2020}.
  Without this term, as in the standard Dicke model, the curvature
  grows with the coupling and the number of emitters and follows the
  superradiant instability, whereas with it the curvature remains of
  order one. In the thermodynamic limit the gauge-invariant model can
  be solved exactly, see~\eqref{eq:dicke-thermo}. The minimal photon
  number then increases monotonically with the coupling, and the gauge
  of minimal photon number is the Jaynes-Cummings gauge
  $\omega_{x}/(\omega_{c}+\omega_{x})$~\cite{stokes_gauge_2019}, at
  all couplings. Our results for finite $N_{s}$ approach this limit
  proportional to $1/N_{s}$, so that the coupling dependence of $\alpha_{\min}$ for
  only a few emitters, which is strongest for the QRM, is a finite-size
  effect.

  The minimum also has an operational meaning. For the QRM it
  equals the photon number of the field based on a projective
  measurement of the dipole, after subtraction of the conditional mean
  field. It thus measures the non-coherence of the conditional field
  states, i.e., their squeezing and non-Gaussianity, and not the
  light-matter entanglement, which itself depends on the gauge. A
  passive measurement of the field noise instead yields a larger value,
  which exceeds the minimum by the squared dipole-field correlator and
  depends on whether the detector responds to the electric or to the
  displacement field. Schemes for detecting virtual photons in
  ultrastrongly coupled
  circuits~\cite{PhysRevResearch.6.013008,giannelliIntegratedConversionPhotodetection2023}
  could be extended to measure this correlator as well, and thus to
  isolate the photn number minimum.

  Since the parabola only requires the displacement form of the gauge
  transformation, it holds for every eigenstate and at finite
  temperature, and several suggestive extensions arise. For matter
  without parity symmetry, such as an asymmetric double well, the truncated
  dipole operator acquires diagonal elements and the model takes the
  form of an asymmetric
  QRM~\cite{li_algebraic_2015,li_addendum_2016,semple_asymptotic_2018,xie_general_2022}.
  Then $G$ is no longer a Pauli matrix, $Q\neq1$ already for a single
  emitter, and the symmetry arguments used for the interpretation of
  the minimum and for $\langle\ad a\rangle_{\min}=N_{\mathrm{cond}}$
  no longer apply. In cavity-coupled lattice models the transformation
  from the Peierls to the dipole gauge is again a displacement
  conditioned on the electronic
  polarization~\cite{dmytruk_gauge_2021}, so that the parabola also
  applies to cavity
  materials~\cite{schlawin_cavity_2022,eckhardt_quantum_2022,schellenberger_almost_2024}.
  There, and in quantum Floquet
  engineering~\cite{eckhardt_quantum_2022,perez-gonzalez_light-matter_2023},
  the conditional non-coherence quantifies the light-matter
  correlations beyond a description of the field by a coherent state.
  Finally, measuring the conditional photon number requires fast
  protocols~\cite{PhysRevResearch.3.023079}, which can be modeled with
  the exact real-time dynamics of the
  QRM~\cite{wolf_exact_2012,wolf_dynamical_2013}.

  \section*{Acknowledgements}
  The authors gratefully acknowledge useful discussions with Daniel
  Braak and Beatriz Pérez González.
  The authors used Anthropic's Claude (up to Opus 5.5) as an
  assistant: it helped to create Fig.~\ref{fig:gauge-orbit-overview},
  supported the writing of parts of the text, helped with parts of the
  code base, and improved some derivations. The authors reviewed all
  of its material and take full responsibility for the content of the
  paper.

  \funding{This project was funded by DFG (German Research Foundation)
  (Project No. 492547816; TRR 360, Project C4)}

  \section*{Data and code availability}
  The code that produces every figure in this paper, together with the
  resulting figures, is openly available at
  \url{https://github.com/ArkaCMTphysics/Gauge_photon}.

  \section*{Competing interests}
  The authors declare no competing interests.

  \nolinenumbers
\end{document}